\documentclass[sigconf]{acmart}
\AtBeginDocument{%
  }

\copyrightyear{2026}
\acmYear{2026}
\setcopyright{cc}
\setcctype{by}
\acmConference[CHI '26]{Proceedings of the 2026 CHI Conference on Human Factors in Computing Systems}{April 13--17, 2026}{Barcelona, Spain}
\acmBooktitle{Proceedings of the 2026 CHI Conference on Human Factors in Computing Systems (CHI '26), April 13--17, 2026, Barcelona, Spain}
\acmPrice{}
\acmDOI{10.1145/3772318.3790720}
\acmISBN{979-8-4007-2278-3/2026/04}

\usepackage{graphicx}        
\usepackage{booktabs}        
\usepackage{array}           
\usepackage{multirow}        
\usepackage{makecell}        
\usepackage{subcaption}      
\usepackage{amsmath}         
\usepackage{float}           

\newcolumntype{Y}{>{\raggedright\arraybackslash}X}

\begin{document}





\title{\textit{Gen-Diaolou}: An Integrated AI-Assisted Interactive System for Diachronic Understanding and Preservation of the Kaiping Diaolou} 



\author{Lei Han}
\orcid{0009-0001-7157-8702}
\affiliation{%
  \institution{Computational Media and Arts, The Hong Kong University of Science and Technology (Guangzhou)}
  \city{Guangzhou}
  \state{Guangdong}
  \country{China}}
\email{lhan229@connect.hkust-gz.edu.cn}

\author{Yi Gao}
\orcid{0009-0007-1267-2495}
\affiliation{%
  \institution{Computational Media and Arts, The Hong Kong University of Science and Technology (Guangzhou)}
  \city{Guangzhou}
  \state{Guangdong}
  \country{China}}
\email{ygao201@connect.hkust-gz.edu.cn}

\author{Xuanchen Lu}
\orcid{0009-0005-8776-8761}
\affiliation{%
  \institution{Computing and Software Technology, Hong Kong Baptist University}
  \city{Hong Kong}
  \country{China}}
\email{22257896@life.hkbu.edu.hk}

\author{Bingyuan Wang}
\affiliation{%
  \institution{Computational Media and Arts, The Hong Kong University of Science and Technology (Guangzhou)}
  \city{Guangzhou}
  \state{Guangdong}
  \country{China}}
\email{bwang667@connect.hkust-gz.edu.cn}

\author{Lujin Zhang}
\orcid{0009-0006-4800-466X}
\affiliation{%
  \institution{Computational Media and Arts, The Hong Kong University of Science and Technology (Guangzhou)}
  \city{Guangzhou}
  \state{Guangdong}
  \country{China}}
\email{lzhang930@connect.hkust-gz.edu.cn}

\author{Zeyu Wang}
\orcid{0000-0001-5374-6330}

\affiliation{%
  \institution{Computational Media and Arts, The Hong Kong University of Science and Technology (Guangzhou)}
  \city{Guangzhou}
  \state{Guangdong}
  \country{China}}
  \email{zeyuwang@hkust-gz.edu.cn}

\author{David Yip}
\authornote{Corresponding author.}
\orcid{0000-0002-1745-4741}
\affiliation{%
  \institution{Computational Media and Arts, The Hong Kong University of Science and Technology (Guangzhou)}
  \city{Guangzhou}
  \state{Guangdong}
  \country{China}}
\email{daveyip@hkust-gz.edu.cn}

\renewcommand{\shortauthors}{}

\begin{abstract}

The Kaiping Diaolou and Villages, a UNESCO World Heritage Site, exemplify hybrid Chinese and Western architecture shaped by migration culture. However, architectural heritage engagement often faces authenticity debates, resource constraints, and limited participatory approaches. This research explores current challenges of leveraging Artificial Intelligence (AI) for architectural heritage, and how AI-assisted interactive systems can foster cultural heritage understanding and preservation awareness. We conducted a formative study (N=14) to uncover empirical insights from heritage stakeholders that inform design. These insights informed the design of \textit{Gen-Diaolou}, an integrated AI-assisted interactive system that supports heritage understanding and preservation. A pilot study ($N$=18) and a museum field study ($N$=26) provided converging evidence suggesting that \textit{Gen-Diaolou} may support visitors’ diachronic understanding and preservation awareness, and together informed design implications for future human--AI collaborative systems for digital cultural heritage engagement. {\color{black}More broadly, this work bridges the research gap between passive heritage systems and unconstrained creative tools in the HCI domain.}

\end{abstract}




\maketitle

\section{Introduction}
Inscribed as a UNESCO World Heritage Site, the Kaiping Diaolou and Villages stand as a unique testimony to the fusion of Chinese and Western architectural traditions, originally funded by overseas emigrants to sustain both defense and dwelling~\cite{unesco_kaiping, zhang2020perceived,batto2006diaolou}. These structures serve as built archives of diaspora ties and the aspirations of modernity~\cite{sun2019place, batto2006diaolou, prott1992cultural}.
However, despite their historical significance, the Diaolou face challenges of conservation and sustainable development, as well as limited participation from local communities~\cite{ryan2011impacts}.
In practice, authenticity debates, resource constraints, and policy misalignment intersect with risks of over-restoration and spectacle-oriented display, making sustained, participatory approaches difficult to realize~\cite{10.1145/3479007, 10.1145/2531602.2531640, wu2015heritage}.

In recent years, within Human-Computer Interaction (HCI), cultural heritage (CH) has emerged as a multifaceted area of study that examines how digital technologies influence human engagement~\cite{muller2025genaichi, museum2024, Fu2024}. As a potential way to enhance engagement, artificial intelligence (AI) has been increasingly applied in CH, enhancing accessibility through multilingual interpretation and digitization. It enables new creative practices such as generative content and interactive storytelling~\cite{zhou2024generative,xu2025review}. Generative artificial intelligence (GenAI) plays an important role in CH engagement by enhancing knowledge retention~\cite{VRflower, 10.1145/3715336.3735808, tao2025aifiligree} and supporting audience co-creation, thereby expanding avenues for participation \cite{wang2025harmonycut, magrisso2018digital, muller2025genaichi, li2024silk, he2025recall}.

Although GenAI accelerates the creative process for non-experts, it entails risks of historical inaccuracy, multimodal instability, and aesthetic homogenization~\cite{childefficacy2024,inbetween2024,lc2023speculative,xu2025review}. Moreover, architectural heritage has largely been overlooked in prior research.
Consequently, the specific challenges associated with deploying GenAI-based systems in real-world scenarios, such as heritage site and museums, and the corresponding strategies for future system design, remain underexplored.

To address this gap, we first conducted a formative study (\textbf{N=14}) that combined preparatory work, including Diaolou data collection and early-stage prototyping, with stakeholder research, comprising expert interviews and two co-design workshop sessions. This process surfaced key cultural themes of the Diaolou, revealed user challenges, and uncovered design opportunities, culminating in five consolidated design goals.

Based on these design goals, we developed \textit{Gen-Diaolou}, an integrated AI-assisted system that supports CH learning and fosters preservation awareness around the Kaiping Diaolou.

We conducted a two-stage evaluation. First, a \textbf{pilot study} (N = 18) at a university assessed usability and workload in a controlled setting and informed subsequent design refinements. Based on these findings, we refined the system and then conducted a museum-based \textbf{field study} (N = 26) to perform a between-subject comparison. 

The results showed that \textit{Gen-Diaolou} effectively supported participants in deepening their knowledge of the Diaolou, both immediately and in delayed measures, and that the GenAI-augmented system can foster awareness of heritage preservation while enabling creative exploration. We discuss design considerations for AI-assisted cultural heritage systems, highlight current limitations, and outline directions for future work.
In summary, this research makes the following contributions:

\begin{itemize}
\item 
A \textbf{formative study}
  that combines preparatory Diaolou data collection and early-stage prototyping with stakeholder research, identifying challenges and requirements through expert interviews and exploring design target through participatory co-design workshops.
\item 
\textbf{\textit{Gen-Diaolou}}, an integrated AI-assisted interactive system, uses a knowledge module and an GenAI module to support theme-based ideation, historically grounded creative exploration, and reflection on CH preservation.
\item 
A two-stage \textbf{empirical study} assessed \textit{Gen-Diaolou} across usability, understanding, and heritage-preservation awareness, yielding design principles for presenting multiple perspectives and implications for future studies on human--AI collaborative heritage learning systems. More broadly, this work bridges the research gap between passive heritage systems and unconstrained creative tools.
\end{itemize}




{\color{black}We use two key constructs throughout the paper.

\textbf{Diachronic understanding} refers to how users connect architectural heritage across time by relating past states, present-day consequences, and future-oriented scenarios. In \textit{Gen-Diaolou}, we scaffold this through an explicit \textit{past--present--future} linkage in the system design, intended to help users explain what changed, why it changed, and how earlier conditions shape current risks and values. We measure diachronic understanding via (i) pre--post gains on time-linked knowledge items and (ii) delayed retention measures that probe temporal reasoning rather than recall of isolated facts.

\textbf{Preservation awareness} refers to an informed, reflective orientation toward safeguarding heritage, including recognition of risks, stakeholder responsibilities, and plausible individual or collective actions to support preservation. In \textit{Gen-Diaolou}, this is supported by risk-oriented and future-planning tasks that prompt users to articulate preservation concerns and possible responses. 
To capture this construct more comprehensively, we adopted a mixed-methods approach: \textit{Study~1} qualitatively explored participants’ reflections on heritage fragility and protection through in-depth interviews, whereas \textit{Study~2} quantitatively assessed the strength of this awareness using a structured Heritage Conservation Awareness scale. Across studies, we observed patterns consistent with gains in participants’ preservation awareness.}

\section{Background and Related Work}

\subsection{Cultural Heritage and the Kaiping Diaolou}

CH refers to the legacy of the physical artifacts and the intangible attributes of a group or society that have been inherited from past generations ~\cite{blake2000defining, vecco2010definition}. Such heritage functions as a critical element in shaping cultural identity and continuity, its essence deeply interwoven with both tangible and intangible dimensions ~\cite{hegedi2023engaging, lvping2021blockchain}.

Historically, the Kaiping Diaolou were constructed during a period of social upheaval and insecurity, serving as fortified residences for overseas Chinese who returned after arduous pursuits of livelihood abroad~\cite{sun2019place}. 
Spanning the late 19th and early 20th centuries, these multi-story towers combined residential, defensive, and aesthetic functions to meet the community's complex needs~\cite{zhang2020perceived, batto2006diaolou,prott1992cultural}. Beyond their physical structures, the Diaolou embody rich socio-cultural narratives~\cite{liu2012analysis} by blending Chinese and Western architectural elements introduced through migration and global exchange. This hybridity illustrates the region’s unique cultural identity and represents a significant chapter in modern Chinese history of adaptation and cross-cultural interaction~\cite{zhang2024architectural}. 

However, the Kaiping Diaolou face considerable challenges in conservation and cultural transmission. Rapid urbanization, shifting demographics, and the pressures of modern development threaten both the integrity and authenticity of these heritage structures~\cite{unesco2006kaiping,zhang2020perceived}. In addition, diminishing community engagement and declining awareness of their historical significance further exacerbate the risks to cultural preservation \cite{zhang2024architectural}. At the same time, there is a lack of effective approaches to engage visitors in understanding and contributing to the preservation of the Diaolou.

\subsection{Leveraging AI in Cultural Heritage Engagement}

AI technologies are increasingly applied in the field of cultural heritage (CH). In HCI domain, researchers have explored user-centered AI designs that enable the public to engage with CH in more immersive and interactive ways, thereby enhancing understanding and raising awareness of preservation~\cite{Fu2024}. Current applications include virtual reconstruction, interactive storytelling, data analysis, and immersive museum experiences~\cite{bordoni2013AIcontribution,ming2024,lg2024multimodal,gameplay2024,narrativegame2020}.

Recent work has used GenAI to design creative tools for CH~\cite{wang2025harmonycut,VRflower,dragondance,10.1145/3613905.3648657}; for example, Tao et al.~\cite{tao2025aifiligree} introduced AIFiligree, an AI-powered framework that generates authentic filigree structures using culturally informed labels and tailored training parameters, improving design efficiency while enhancing cultural communication.

The implementation of AI-augmented systems in CH offers users an accessible and immersive interactive experience, facilitating \textit{legitimate peripheral participation} \cite{1991situated}, through which users can gain knowledge within museums or heritage sites and develop practices related to CH \cite{museum2024,Fu2024}. Previous studies have demonstrated that GenAI-based workflows can enhance users' understanding of CH, encourage reflection, and foster deeper emotional connections, thereby raising their awareness of CH preservation \cite{liu2024hyper, wen2024AIGC,Fu2024,he2025recall}.

Despite rapid advances, current AI practices in CH still face notable limitations in balancing historical accuracy with algorithmic creativity~\cite{oppenlaender2025artworks, 10.1145/3746027.3755670, tohidi2006getting,childefficacy2024}. The core challenge in deploying GenAI for educational purposes in CH lies in mitigating the inherent risk of hallucination and cultural misrepresentation~\cite{boiano2024ethical}.  In CH contexts, such systems often struggle to capture linguistic nuance, represent local artistic and architectural forms, and accurately reflect social and cultural diversity~\cite{inbetween2024,10.1145/3678698.3687200,xu2025review}. These shortcomings can also lead to ethically and culturally problematic outputs that distort historical facts or conflict with prevailing social norms~\cite{zhang2020perceived,inbetween2024}.

For instance, He et al. found that using GenAI without customization led to missing cultural features and biases, especially for unfamiliar sites~\cite{he2025recall}. To mitigate these challenges, recent studies have explored a range of approaches, including domain-specific model training, enhancing user control~\cite{wang2025harmonycut}, and incorporating contextual or cultural knowledge into generative workflows~\cite{ferretti2025ai}.  Notably, Ferretti~\cite{ferretti2025ai} leveraged Retrieval-Augmented Generation (RAG) to enable heritage-based dynamic storytelling in an educational context. However, research on the systematic embedding of such customized AI systems directly into museum scenarios for public engagement remains scarce.

To address this problem, we use the Kaiping Diaolou as a case to examine the challenges people face when using AI tools for cultural-heritage image creation. Based on these insights, we derive design goals and develop an integrated AI system to enhance users’ understanding and foster preservation awareness.

\subsection{Facilitating Diachronic Narrative Creation through Generative AI}

Heritage is not merely a physical entity such as an object, site, or event, but a dynamic cultural and social narrative process~\cite{Fu2024}. However, the transmission of intangible historical elements to the public remains challenging~\cite{narrativegame2020}. Existing approaches to CH engagement range from traditional on-site visits—often limited by time and space—to digital alternatives like digital museums~\cite{museum2024}, online lectures~\cite{kraybill2015going}, interactive games~\cite{luo2025generative}, and VR experiences~\cite{li2024museumvr,museum2024}. Yet, these methods largely convey knowledge from a third-person perspective, neglecting users’ emotional connection with artifacts and limiting the sustainability of affective immersion~\cite{gameplay2024,Cai2024}, thereby failing to capture the deeper, intangible relationships that bind people to CH~\cite{Fu2024}.

Given that narrative is central to human experience and a fundamental mechanism for meaning-making~\cite{madej2003towards,antony2024story}, it serves as an effective tool for communicating complex concepts. This capacity is amplified in interactive digital narratives~\cite{atmaja2022when}, which facilitate a paradigm shift from passive consumption to active participation. In this context, audiences engage with personalized content responsive to their choices and inputs, which in turn fosters their interest and emotional bonds towards CH~\cite{hashim2019narrative,li2023applications,antony2024story}. 

For instance, Antony and Huang introduced ID.8~\cite{antony2024story}, allowing customization in the co-creation of visual stories. A recent study by Trichopoulos et al.~\cite{Trichopoulos2025chatbot} integrated LLM-based chatbots into diverse museum settings, demonstrating their capacity to provide personalized narrative experiences and significantly enhance visitor engagement. Studies indicate that contextualized stories and scenarios can bridge personal experiences with cultural and social issues, providing a more intuitive understanding of potential futures, which in turn contribute to preservation awareness~\cite{ruller2022speculative,cemeteries2022design}.

Emphasizing the temporal significance of CH and establishing a narrative that spans from the past to the present and into the future has been shown to enhance users' emotional connection with CH~\cite{Fu2024,he2025recall}. A notable example is the work of Fu et al.~\cite{Fu2024}, which employed MidJourney, a text-to-image tool, to prompt reflections on the future of CH. Their study demonstrated that GenAI-assisted co-creation experiences can foster personal narratives and critical reflection. Nevertheless, such efforts remain limited, as most studies rely on existing generative tools~\cite{Fu2024,he2025recall}, and focus largely on the generation process or qualitative feedback~\cite{childefficacy2024,lc2023speculative}, leaving a gap in experimental studies that verify the link between participatory storytelling and enhanced understanding and preservation awareness. 

To address this gap, our work aims to employ GenAI to transform the static architecture and knowledge of Diaolou into dynamic cultural narratives, thereby fostering deeper understanding and preservation awareness.

\section{Formative Study} \label{FS}

The formative study combined preparatory work (Section~\ref{Preparatory}), expert interviews (Section~\ref{Expert}) and two co-design workshops (Section~\ref{Workshop}). The aim was to elicit first-hand insights and actionable ideas. These were distilled into design insights and translated into design goals (Section ~\ref{Design Goals}). 

{\color{black}This study was reviewed and approved by the ethics committee at the first author’s institution, all participants provided written informed consent, workshop participants received 60 CNY (approximately 8.4 USD), and all audio recordings, sketches, and generated images were anonymized, stored under restricted access, and used solely for research. Identifiable details were removed during transcription and analysis to protect privacy.}

\begin{figure}
    \centering
    \includegraphics[width=1\linewidth]{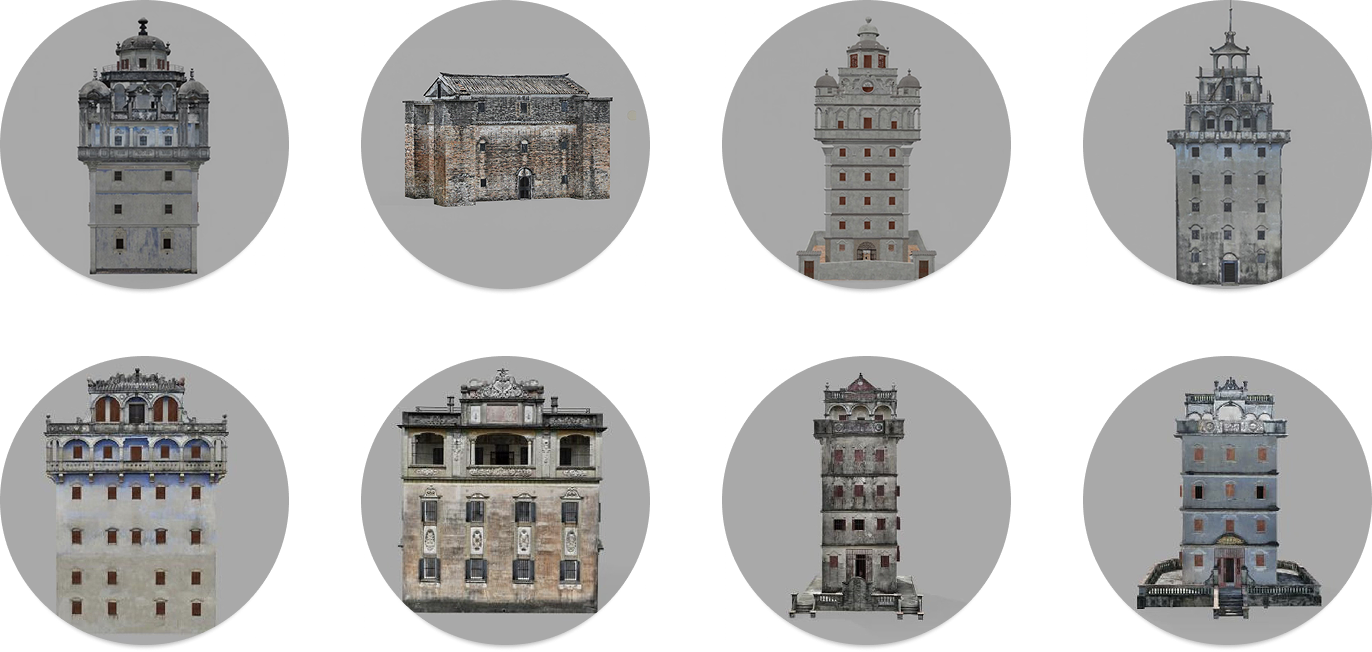}
    \caption{Examples of representative Kaiping Diaolou sites included in our data collection.}
    \Description{Examples of representative Kaiping Diaolou sites included in our data collection.}
    \label{fig:diaolouphoto}
\end{figure}

\subsection{Preparatory Work}\label{Preparatory}

\subsubsection{Diaolou Data Collection}

We conducted archival research and data curation on the Kaiping Diaolou, drawing on archival sources, field documentation across major clusters, and prior scholarship \cite{zhang2020perceived, batto2006diaolou, yuxin2023research, chiang2021landscapes}. 
The Diaolou image collection criteria were: (i) representativeness across major clusters (\textit{e.g., Sannienli Village, Zili Village 
and the Fang Clan Diaolou, Majianglong Village cluster, and Jinjiangli Village}), functional types, and display representative style features (Indo–British, Baroque, Neoclassical, eclectic); (ii) accessibility for documentation and public visitation (availability of archival and photographic materials as well as usage permissions); and (iii) photographic completeness and conservation status. 

The image corpus used for coding consisted of high-resolution photographs obtained from publicly accessible platforms operated by local CH and government authorities. We then selected ten representative Diaolou sites using stratified purposeful sampling to maximise coverage and recognisability (see Figure~\ref{fig:diaolouphoto}). Their attributes were organised into three categories: functions, stylistic idioms and structural components\footnote{Further details on classification examples are summarized in supplementary materials}.

\subsubsection{Prototype Development}

We built an early prototype to let users explore Kaiping Diaolou visual themes with GenAI. It supports text-to-image and image-to-image generation from natural-language prompts, using a transparent, node-based workflow in \textit{ComfyUI}\footnote{ComfyUI, Available at \url{https://www.comfy.org/zh-cn/}}. 
A curated corpus of Diaolou-classified images is included so participants can cite or import references during exploration.

\subsection{Expert Interview}\label{Expert}

\subsubsection{Participants and Procedure} 
We interviewed two senior professionals: \textbf{E1}, a heritage scholar specializing in the conservation of the Kaiping Diaolou, and \textbf{E2}, a local museum director with extensive experience in organizing participatory activities for CH engagement. Both had over 20 years of professional experience\footnote{Further details on participant demographics are in Appendix A}. Two semi-structured interviews (approximately 50 minutes each) were conducted via online video conferencing. With participant consent, all sessions were audio-recorded and transcribed. 

Two authors co-coded an initial subset to develop and align a codebook based on thematic analysis \cite{clarke2017thematic}, which  was then applied to the remaining data; discrepancies were resolved through discussion with periodic peer debriefs. The interview protocol is provided in supplementary materials. 
As part of the interviews, we also presented images generated from the early-stage prototype to the experts to solicit feedback on cultural authenticity and design relevance.

\subsubsection{Design Insights (DI1–DI3): Challenges and Requirements Identified in Expert Interviews}\label{expert insights}

\paragraph{\textbf{DI1: The Need for Historical Accuracy in GenAI Outputs}}
Expert interviews revealed a dual challenge in preserving and disseminating the Kaiping Diaolou and Villages: reconciling innovation with historical authenticity ($E1$). On the preservation side, only a few Diaolou are formally protected, while over-commercialization and historically inaccurate representations undermine authenticity and distort history. Regarding dissemination, experts noted frequent inaccuracies generated by GenAI tools, stressing that any application in this domain must prioritize CH accuracy to prevent exacerbating these distortions ($E1$).


\paragraph{\textbf{DI2: Limitations in Visitor Experience and the Potential of Immersive Narrative}}
Regarding visitor experience, experts highlighted key limitations of the current touring model. Exhibitions on Diaolou history remain confined to museums, while the sites themselves are scattered across Kaiping villages. At each site, there is a lack of interactive forms of interpretation and visitor engagement, leaving visitors to passively observe the architectural exteriors and making it difficult for them to gain a deeper understanding of the profound cultural values these structures embody ($E1, E2$). To address this limitation, experts advocated exploring immersive experiences that integrate historical scene reconstruction and authentic character storytelling, while ensuring historical accuracy ($E1, E2$). 



\paragraph{\textbf{DI3: Challenges of Scaling and Disseminating Diaolou Knowledge}}

In current museum practices, various gamified elements have been introduced, integrating on-site museum experiences with educational programs on Diaolou culture developed in collaboration with domain experts, including creative activities such as themed hand drawings. However, large-scale promotion and dissemination remain limited by high human and resource demands and the short duration of visits, which significantly constrain the scalability of immersive cultural experiences and broader public access ($E2$).

\begin{figure*}[ht]
\centering
\includegraphics [width=\textwidth]{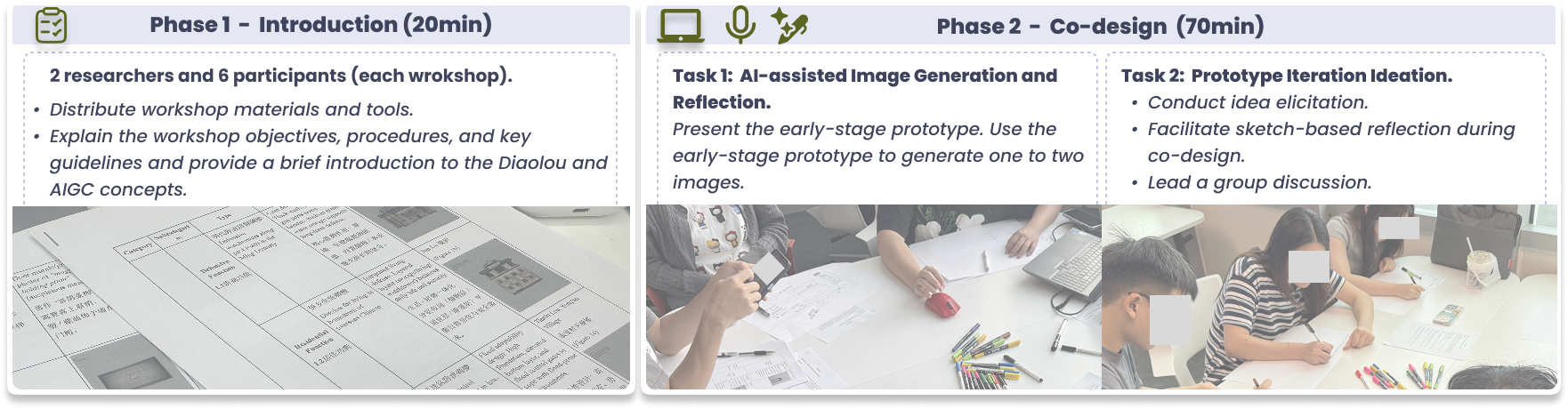} 
\caption{Co-design workshop procedure. The session included Phase~1, Introduction (20 minutes), and Phase~2, Co-design (70 minutes). Participants engaged in two tasks: Task~1, AI-assisted Image Generation and Reflection; and Task~2, Prototype Iteration and Ideation.} 
\Description{Co-design workshop procedure. The session included Phase~1, Introduction (20 minutes), and Phase~2, Co-design (70 minutes). Participants engaged in two tasks: Task~1, AI-assisted Image Generation and Reflection; and Task~2, Prototype Iteration and Ideation.}
\label{5} 
\end{figure*}

\subsection{Co-design Workshop} \label{Workshop}

We then conducted a user co-design workshop followed user centered design principles \cite{abras2004user} to collect feedback and iteratively refine the prototype, ensuring that the experience aligned with both educational and engagement objectives. 

\subsubsection{Participants and Procedure}

We recruited 12 university students (6 female, 6 male; $M=23.33$, $SD=3.04$) via student group chats and social media reposts for two on-campus workshops (Session~\textbf{WS-A} and Session~\textbf{WS-B}; 6 participants each)\footnote{Further details on participant demographics are in Appendix A}. Each session included one participant with a design background and one with technical development experience to balance technical, creative, and user perspectives during co-design activities. 

We used an early-stage prototype and a classification sheet of Diaolou references (refer supplementary materials). We provided A3 sketch sheets, sticky notes, and markers for participants to externalize their ideas and collaboratively develop design concepts. Each 90-minute face-to-face workshop on campus comprised two phases (see Figure~\ref{5}):

\textit{\textbf{Phase 1: Introduction (20 min).}}
After consent and a short demographics survey, participants received a standardized briefing:
(i) the historical and cultural significance of the Kaiping Diaolou;
(ii) basic prototype concepts; and
(iii) the provided materials and tools.

\textit{\textbf{Phase 2: Co-design (70 min).}}
Participants completed two tasks.
\emph{Task i: AI-assisted image generation and reflection.} Participants used the prototype to generate one to two images of the Kaiping Diaolou, experimenting with text-to-image and image-to-image workflows. A group discussion followed, in which participants presented their outputs and exchanged views on \textit{encountered challenges, thematic preferences, and suggestions}.
\emph{Task ii: Prototype ideation.} Building on the prior discussion, participants outlined potential user interfaces, system functions, and additional features, which they then shared to elicit feedback and enable comparison. This step was intended to make the earlier discussion more concrete and actionable.

\subsubsection{Design Insights (DI4--DI6): Design Opportunities from Co-design Workshop}\label{Design insights}

\paragraph{\textbf{DI4: Enhancing Engagement through First-Person Narrative Perspectives}}  

Participants suggested adding first-person narratives to make the learning experience more immersive 
. For example, $C7$ noted that being guided by a historical persona made the process feel less monotonous and more personal. Similarly, $C10$ and $C12$ suggested introducing an persona that could ask them related questions. These reflections highlight the need to embed narrative-driven personas and storylines to sustain attention and enhance cultural learning.

\paragraph{\textbf{DI5: Balancing Authenticity with Creative Freedom}}  
Participants (e.g., $C1, C3, C5, C7, C8, C11, C12$) emphasized that creativity should \emph{not} come at the expense of the Diaolou’s architectural body. Even when reference images were provided, the model sometimes weakened or omitted distinctive features, yielding implausible or uncanny façades. Others ($C2, C6, C9$) reported frustration that iterative prompt refinements had little effect when outputs contradicted historical facts. Together these concerns reveal limits in both authenticity and user control, underscoring the need for reference locking, heritage-informed constraints, and more transparent editing mechanisms beyond text-only prompts.


\paragraph{\textbf{DI6: Expanding Content Diversity with Scaffolded Support}}  

Participants called for broader creative scope and stronger system reference support. They wanted interior options (e.g., magpie, plum, pine–crane, \emph{fu}-characters) rather than façades alone ($C4, C7, C10, C11$), but also admitted lacking detailed knowledge of architectural elements (e.g., window decorations, structural components). Coupled with difficulties in formulating precise prompts, this suggests the need for modular exemplars, motif libraries, and structured scaffolds that help users translate intent into effective inputs.

\subsection{Design Goals}\label{Design Goals}

Drawing on the formative study, we derived five design goals for an integrated AI-assisted interactive system that supports engagement with the history of the Diaolou and fosters awareness of heritage preservation:
\begin{itemize}
   
  \item \textbf{DG1. Ensure historical accuracy in generative outputs.}
The system should provide mechanisms to minimize factually incorrect or misleading representations of Diaolou and related heritage content (\textit{DI1}).
  \item \textbf{DG2. Enable immersive, narrative-driven engagement.}
The system should support first-person and story-based interactions that enrich user experience and foster empathy with heritage narratives (\textit{DI2, DI4}).
  \item \textbf{DG3. Balance authenticity with creative freedom.}
The design should safeguard cultural authenticity while allowing users space for creative exploration and personalization (\textit{DI5}).
  \item \textbf{DG4. Support scalable knowledge dissemination.}
The system should facilitate the communication of Diaolou-related knowledge to diverse audiences beyond the local context (\textit{DI3}).
  \item \textbf{DG5. Provide diverse content and scaffolded guidance}
The system should offer a rich set of content resources alongside scaffolding strategies that help users explore, reflect, and co-create meaningfully (\textit{DI6}).

\end{itemize}

\section{System Design}~\label{system}


Based on the design goals (Section~\ref{Design Goals}), we developed \textit{Gen-Diaolou}, an integrated AI-assisted interactive system to support CH learning and foster preservation awareness through a \textit{learn--then--create} flow. The system comprises a \textit{Knowledge Module} (see Figure~\ref{fig:genai-ui}) and a \textit{GenAI Module} (see Figure~\ref{ui1.1}). Global controls include a language switch (Chinese/English) and accessibility options for larger text and high-contrast display.

\subsection{Design Objectives}
The system architecture enhances the accessibility of Diaolou heritage knowledge, enabling dissemination beyond geographic constraints to a broader audience (\textit{DG4, DI2}). The user journey is framed as a narrative experience, guided by a historical persona that orients visitors and deepens immersion (\textit{DG2}). Within the \textit{GenAI Module}, the creative process begins with structured selections that surface diverse content and provide scaffolded guidance before users articulate free-form ideas (\textit{DG5}). The module then applies output constraints and checking mechanisms that promote historically grounded yet imaginative outputs, helping the system balance accuracy and creative freedom (\textit{DG1}, \textit{DG3}).

\subsection{System Components}\label{Core}

\begin{figure}[t]
  \centering
  \includegraphics[width=\columnwidth]{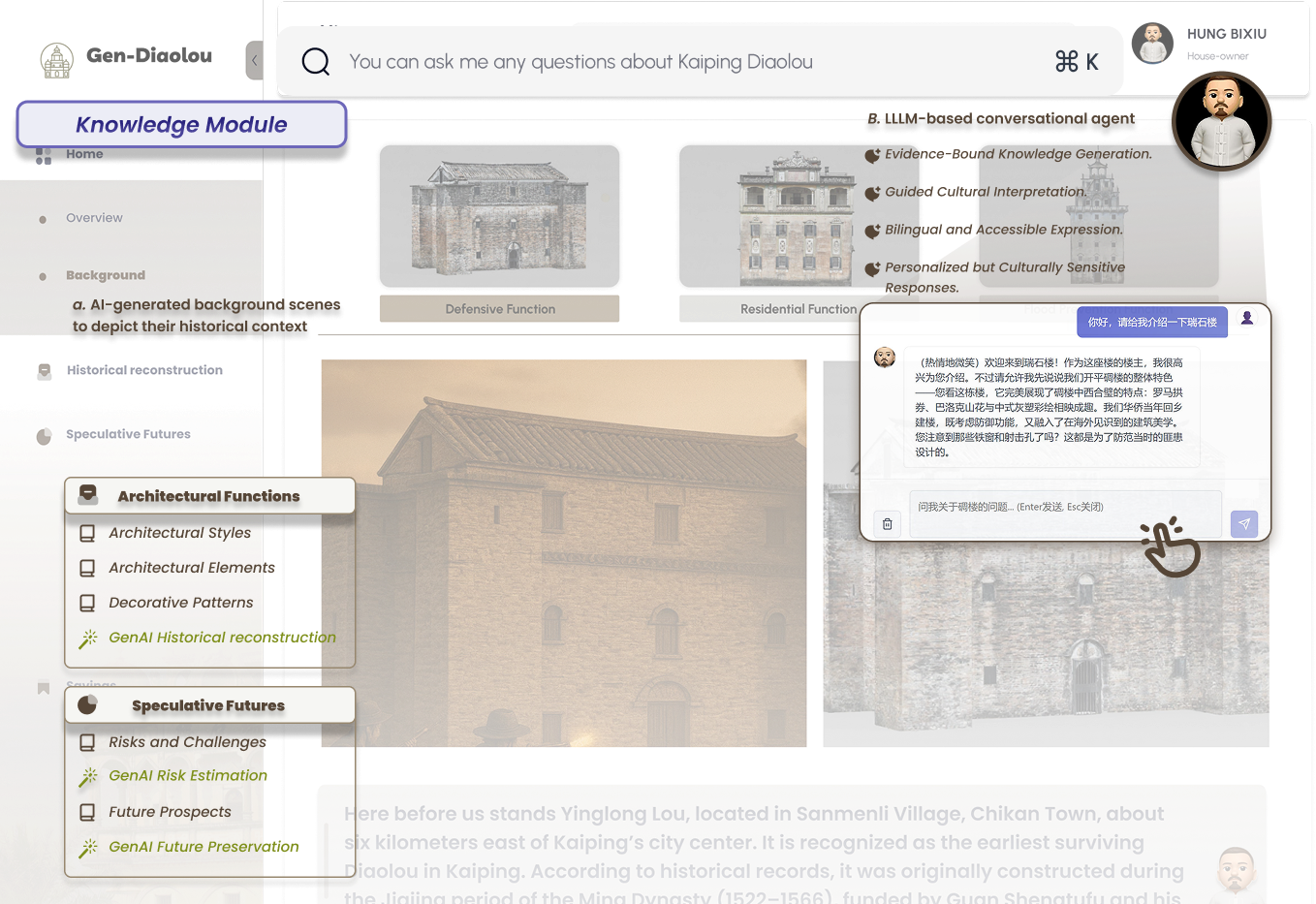}
  \caption{UI of the \textit{Gen-Diaolou} Knowledge Module.}
  \Description{User interface of the Gen-Diaolou Knowledge Module.}
  \label{ui1.1}
\end{figure}

\subsubsection{Knowledge Module}

The module supports preparatory learning through three sub-sections: \textit{Background}, \textit{Historical Reconstruction}, and \textit{Speculative Futures}, which together provide narrative- and taxonomy-based access to curated materials derived from our prior work (Section~\ref{Preparatory}) (\textbf{DG1}).  

First, \textit{Background} offers an overview of the Kaiping Diaolou, including a map of their geographic distribution, and an interactive storytelling interface that uses AI-generated background scenes to depict their historical context (see Figure~\ref{ui1.1}). Here, \textit{Huang Bixiu}, builder of \textit{Ruishi Lou} \cite{batto2006diaolou}, appears as an LLM-based conversational agent who narrates each section and introduces the origins and social context of the Diaolou  (\textbf{DG2}).

Secondly, \textit{Historical Reconstruction} provides modular learning resources via a taxonomy navigator, including classifications of architectural functions and styles, as well as interior decorative motifs, to structure key concepts (see Figure~\ref{ui1.1}). Finally, \textit{Speculative Futures} introduces major preservation challenges currently facing Diaolou heritage and situates subsequent creative exploration within a realistic heritage context, grounding later design work in actual preservation issues (\textbf{DG4}).


\begin{figure*}[h]
  \centering
  \includegraphics[width=\textwidth]{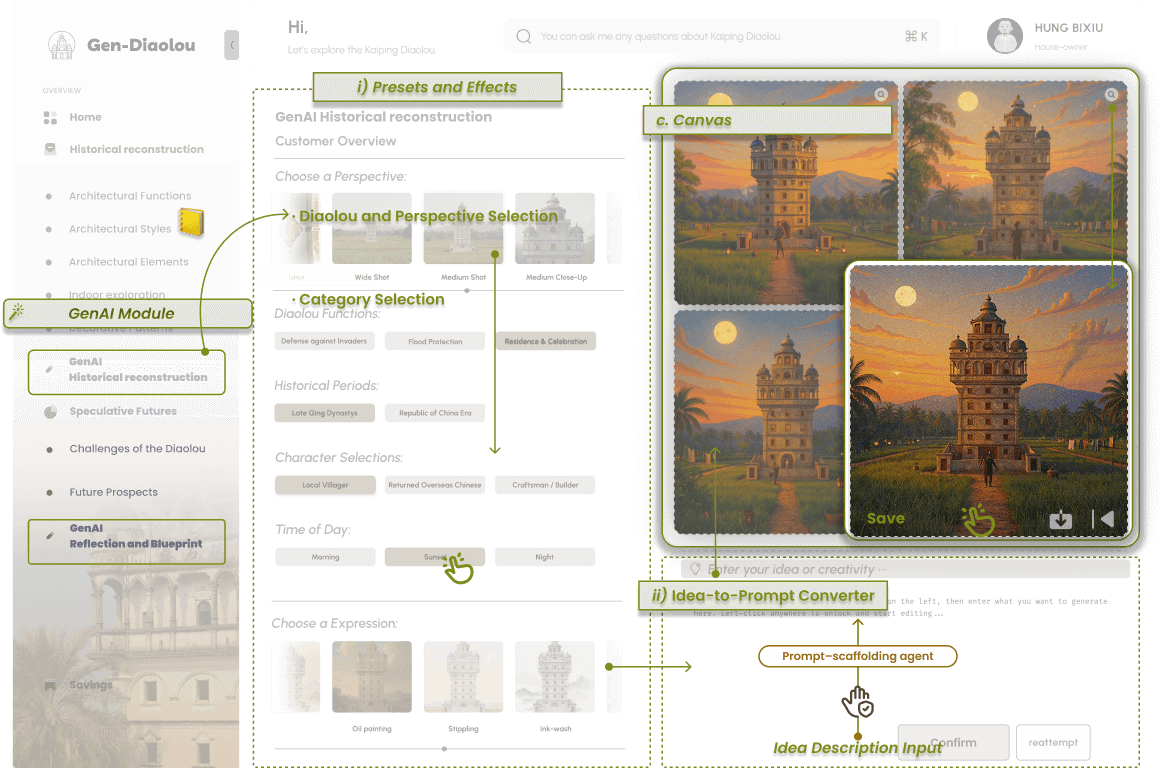}
  \caption{UI of the \textit{Gen-Diaolou} GenAI Module: (i) presets and effects, (ii) an idea-to-prompt converter with LLM scaffolding, and (iii) a canvas with refine and save controls.}
  \Description{The figure shows the user interface of the Gen-Diaolou GenAI Module. At the top, users can select preset Diaolou exemplars and visual effects. In the middle, an idea-to-prompt panel uses an LLM to scaffold and expand the user’s brief idea into a detailed, validated prompt. On the right, a two-by-two image canvas displays four generated scenes, with controls to refine each image and buttons to save selected results.
}
  \label{fig:genai-ui}
\end{figure*}

\begin{figure*}[h]
  \centering
  \includegraphics[width=\textwidth]{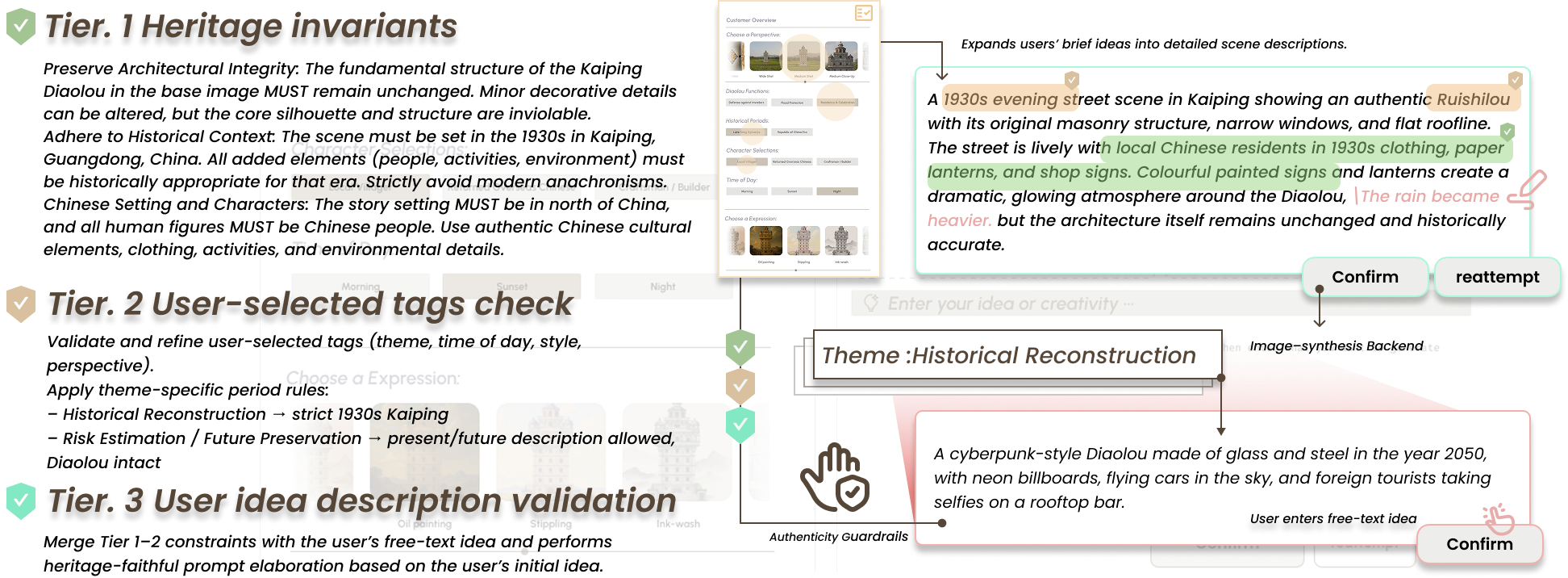}
    \caption{Detailed example of the authenticity guardrails workflow for \textit{Historical Reconstruction} in the \textit{GenAI Module}.}
  \Description{Detailed example of the authenticity guardrails workflow for Historical Reconstruction in the GenAI Module.}
  \label{fig:genai-workflow}
\end{figure*}

\subsubsection{GenAI Module}\label{AIm}

The module is implemented across three creative sub-sections—\textit{Historical Reconstruction}, \textit{Risk Estimation}, and \textit{Future Preservation} (\textbf{DG5})—which share a common interaction flow, with the user interface shown in Figure~\ref{fig:genai-ui}.

First, users select from presets and effect options tailored to the current sub-section. These controls surface recommended content, contextual knowledge, and explanations (e.g., different perspectives, historical periods, or visual emphases), helping users make informed choices. 
Next, users provide a text-based idea description to the \emph{prompt–scaffolding agent}. The agent retrieves the corresponding category descriptions from a curated heritage knowledge base and composes a structured prompt, which is shown back to the user to check whether it matches their original intent. 

During this process, \emph{Authenticity Guardrails} adjust or constrain the prompt to better align with key historical facts (Section~\ref{point1}). Upon user confirmation, the finalized prompt is sent to the \emph{image–synthesis backend} to generate the image set (Section~\ref{point3}).


\subsection{Design Features}

\subsubsection{Persona Design.}

To support inquiry-based learning and immersive engagement (\textbf{DG2}), the system features an LLM-based conversational agent: \textit{Huang Bixiu}, the historical builder of \textit{Ruishi Lou} \cite{batto2006diaolou}. In the \textit{Knowledge Module}, \textit{Huang Bixiu} narrates sections and introduces the origins and social context of the Diaolou. In the \textit{GenAI Module}, he responds to user questions about the Kaiping Diaolou, allowing users to seek clarification through natural dialogue complemented by conversational guidance grounded in curated heritage knowledge.

\subsubsection{Authenticity Guardrails.}\label{point1}

We present a three-tier authenticity guardrail flow shown in Figure~\ref{fig:genai-workflow}, integrated into a prompt-scaffolding agent, that systematically elaborates users’ brief ideas into detailed scene descriptions while keeping outputs historically grounded and culturally appropriate to the Kaiping Diaolou context (\textbf{DG1}, \textbf{DG3}).

\textbf{Tier~1 — Heritage invariants.} Non-negotiable rules preserve the Diaolou’s structural identity (architectural form, proportions, façade details, window positions, roofline). Only the surrounding environment may be modified; all scenes must be set in Kaiping, Guangdong, China, and all human figures and cultural elements must remain consistent with Chinese cultural heritage.

\textbf{Tier~2 — Validation of user-selected tags.} Preset tags (e.g., Diaolou exemplars, viewpoint, time of day, season; are treated as hard requirements and directly guide prompt assembly (see Table~\ref{ttier2} in Appendix).

\textbf{Tier~3 — User idea validation.} The user’s free-text idea description is checked and normalized to comply with Tiers~1–2 before being incorporated. If the idea description conflicts with the tags, the tags take precedence. If any input violates theme-specific rules, it is automatically normalized according to Tier~1 constraints; non-conforming descriptions are rewritten or removed.

The guardrails apply different validation strategies across different task themes. In \textit{Historical Reconstruction}, the framework strictly requires all scene elements (people, activities, clothing, objects, environment) to conform to the 1930s period in Kaiping, applying conservative validation to enforce period-correct details. In \textit{Risk Estimation} and \textit{Future Preservation}, the framework relaxes the temporal requirement to allow present or future scenarios while preserving the recognizable Diaolou form, permitting greater divergence in scene composition.

Lastly, the validated prompt is presented for user review and editing prior to image generation; any edits trigger revalidation under the same hierarchy. Unlike static templates, these guardrails \emph{adapt constraint strength to the task context}, operationalizing domain knowledge into structured, adaptive rules that balance accuracy with creative freedom and turn raw user ideas into consistent, heritage-faithful prompts (see Appendix~\ref{appendix:guardrails} for complete specifications).

\subsubsection{Image–synthesis Backend.}\label{point3}

The image–synthesis backend combines \textit{ComfyUI}\footnote{ComfyUI. Available at~\url{https://www.comfy.org/zh-cn/}} with the FLUX.1 Kontext Pro model
for image-to-image generation. \textit{ComfyUI} orchestrates the generation workflow through a node-based interface, providing programmatic control over the pipeline and allowing us to inject authenticity constraints derived from the guardrails. 

Within this workflow, the \textit{FLUX.1 Kontext Pro} model acts as the core synthesis engine: given a guardrail-constrained prompt and a Diaolou base rendering, it preserves the global structural layout while enabling targeted, local edits, ultimately producing the final images. This controllable backend is invoked whenever the \textit{GenAI Module} confirms a prompt, supporting both historically grounded reconstructions and the speculative risk and preservation scenarios described above.

\subsection{Implementation Details}

Our system adopts a decoupled client–server architecture, with a Python backend built on FastAPI~\cite{fastapi} and a frontend implemented in Vue 3 with \textit{Vite}. The frontend is packaged as a Progressive Web App (PWA)~\cite{google-pwa,vite-plugin-pwa}, using Service Workers~\cite{service-worker} to cache the application shell and static educational content for offline access while keeping network connectivity for GenAI features. 

We chose a PWA over native mobile applications to balance cross-platform accessibility (browser-based usage) with app-like capabilities (home screen installation, offline mode), which is critical in museum and on-site heritage settings with potentially unstable Wi-Fi. The client and server communicate via RESTful APIs, and the backend exposes a proxy endpoint for serving generated images to mitigate Cross-Origin Resource Sharing (CORS) constraints~\cite{whatwg-fetch}. For LLM-based features, the system integrates DeepSeek v3~\footnote{DeepSeek v3. Available at~\url{https://github.com/deepseek-ai/DeepSeek-V3}} to power the conversational agent and the prompt-scaffolding agent within the authenticity guardrails.

\l \label{US}

We conducted a two-stage empirical evaluation study. First, we conducted a \textbf{pilot study} (\(N = 18\)) at a university to assess usability and workload in a controlled setting and to derive actionable design improvements (Section~\ref{S-1}). 

Based on these findings, we refined the system and then ran a museum \textbf{field study} (\(N = 26\)) with a more diverse participant pool to begin assessing external validity and to evaluate the incremental effect of the \textit{GenAI Module} by comparing a \textbf{Base} condition (no \textit{GenAI Module}) to a \textbf{Learn+GenAI} condition (\textit{knowledge Module} + \textit{GenAI Module}) (Section~\ref{S-2}).

{\color{black}This study was reviewed and approved by the ethics committee at the first author’s institution. All participants provided informed consent and could withdraw at any time without penalty.} We collected only study-relevant data (questionnaires, interaction logs, sketches, interviews); personally identifiable information was not stored with research data. Audio/visual materials were used solely for analysis and de-identified during transcription and reporting. Audio was transcribed verbatim (Chinese) using a commercial \textit{ASR system (iFLYTEK)}\footnote{iFLYTEK. Available at~\url{https://www.iflyrec.com/zhuanwenzi.html} (accessed May~2025).}, thematically coded using a bottom-up approach \cite{clarke2017thematic}, and translated into English with back-translation by two bilingual researchers to ensure conceptual equivalence.


\begin{figure*}[ht]
  \centering
  \includegraphics[width=\textwidth]{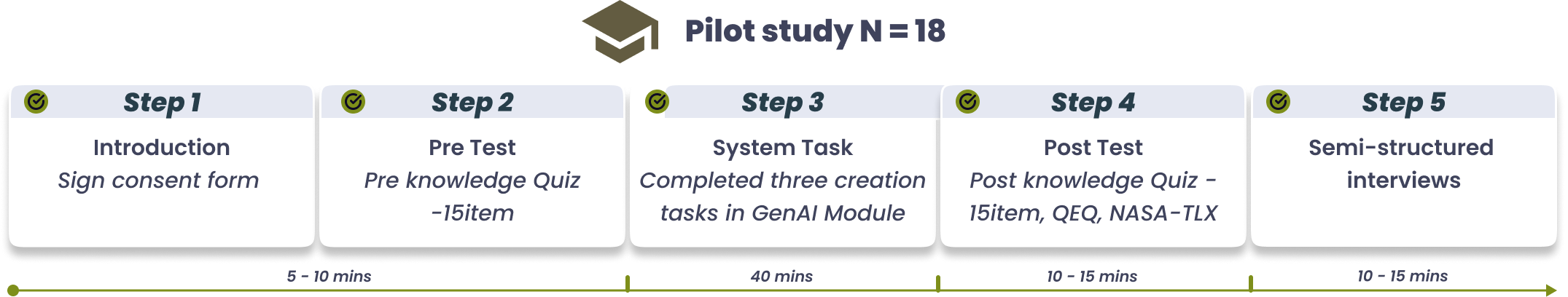}
  \caption{Pilot study procedure diagram.}
  \Description{Pilot study procedure diagram.}
  \label{procedurefig}
\end{figure*}

\section{Study 1: Pilot Study}\label{S-1}

We conducted a pilot study (N=18) to gather feedback and iteratively refine the prototype for alignment with learning and cultural-heritage preservation objectives, and to assess subjective workload (\textit{NASA-TLX}) and user experience (\textit{UEQ}). {\color{black}Participants received 60 CNY (approximately USD 8.4) as compensation.}

\subsection{Participants and Procedure}

A total of 18 participants took part in this pilot study (7 female, 11 male; $M = 24.17$, $SD = 3.81$)\footnote{Further details on participant demographics are in Appendix A}, recruited via social media postings at a university and providing self-reported demographic information. Participants met the criteria: 18 years or older, basic digital literacy, and an interest in CH.

\begin{figure*}
  \centering
  \includegraphics[width=\textwidth]{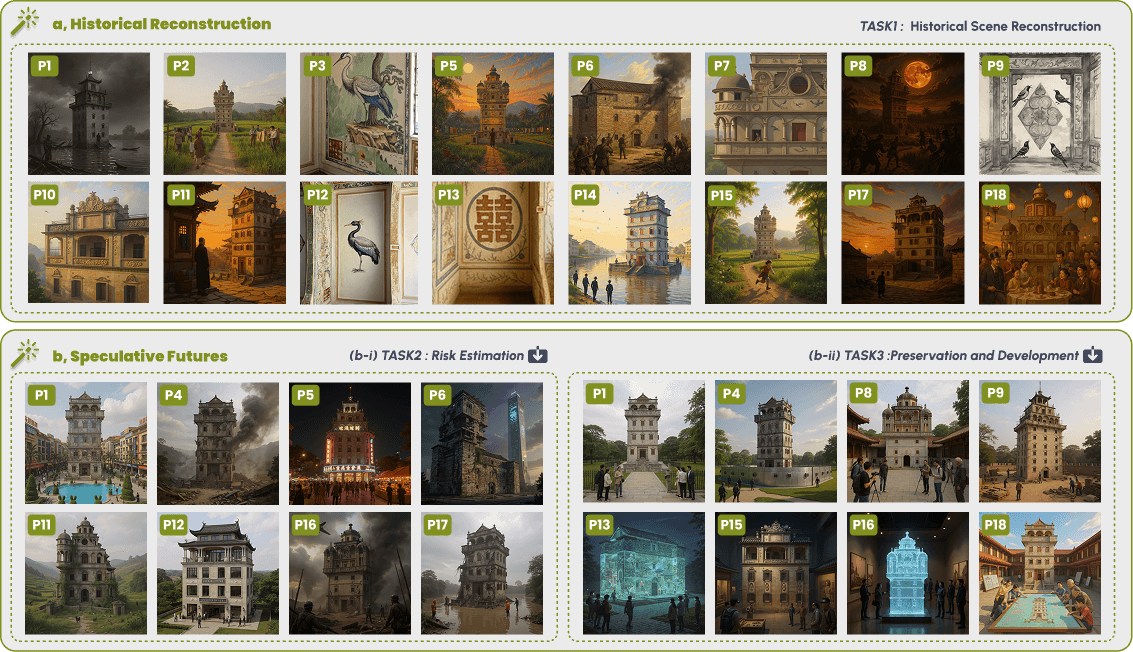}
  \caption{The following illustrations exemplify participant-generated images produced with \textit{Gen-Diaolou} during the user study, and illustrate two themes: a. \textit{Historical reconstruction} and b. \textit{Speculative Futures}.}
  \Description{The following illustrations exemplify participant-generated images produced with Gen-Diaolou during the user study, and illustrate two themes: a. \textit{Historical reconstruction} and b. Speculative Futures.}
  \label{cover}
\end{figure*}

The pilot study procedure is shown in Figure~\ref{procedurefig}. After providing informed consent and demographic information, participants first completed a 15-item quiz (maximum score = 15) to measure their initial knowledge about the Kaiping Diaolou.
Next, participants freely explored the system, moving from learning to creation. They consulted the \textit{Knowledge Module}’s taxonomy navigator as needed, then completed three content creation tasks (see Figure \ref{cover}) in \textit{GenAI Module}, iteratively producing 2×2 image grids until at least one satisfactory result per task. 
After completing all tasks, participants took a post-study knowledge quiz and filled out UEQ and NASA-TLX. Finally, semi-structured interviews were conducted by the first author to collect reflections on system experience.

\subsection{Evaluation Dimensions}\label{q}

We assessed \textbf{learning outcomes} using a blueprint-based 15-item pre/post quiz derived from authoritative sources on the Kaiping Diaolou. Details of the quiz blueprint and item pool are reported in supplementary material.
\textbf{User experience} was measured using the \textit{User Experience Questionnaire} (UEQ) \cite{laugwitz2008construction} on a 7-point Likert scale.
\textbf{System workload} was measured with the \textit{NASA Task Load Index} (NASA-TLX) \cite{hart2006nasa}: mental demand, physical demand, temporal demand, performance, effort, and frustration (see Table \ref{tab:ueq_nasa_mapping} in Appendix). In our study, responses were collected on a 7-point Likert scale (1 = best, 7 = worst).
\textbf{Task performance} was quantified from system logs (task ID, inputs/parameters, iteration counts, saved images) as iterations-to-accept, total iterations, and number of saved images.
\textbf{Qualitative feedback} came from 10–20-minute semi-structured interviews on usability, creativity support, and heritage engagement. The interview protocol is provided in supplementary materials.

\subsection{Findings}

\subsubsection{Impact of Learning Outcomes}

\begin{figure}[htbp]
    \centering
\includegraphics[width=1\linewidth]{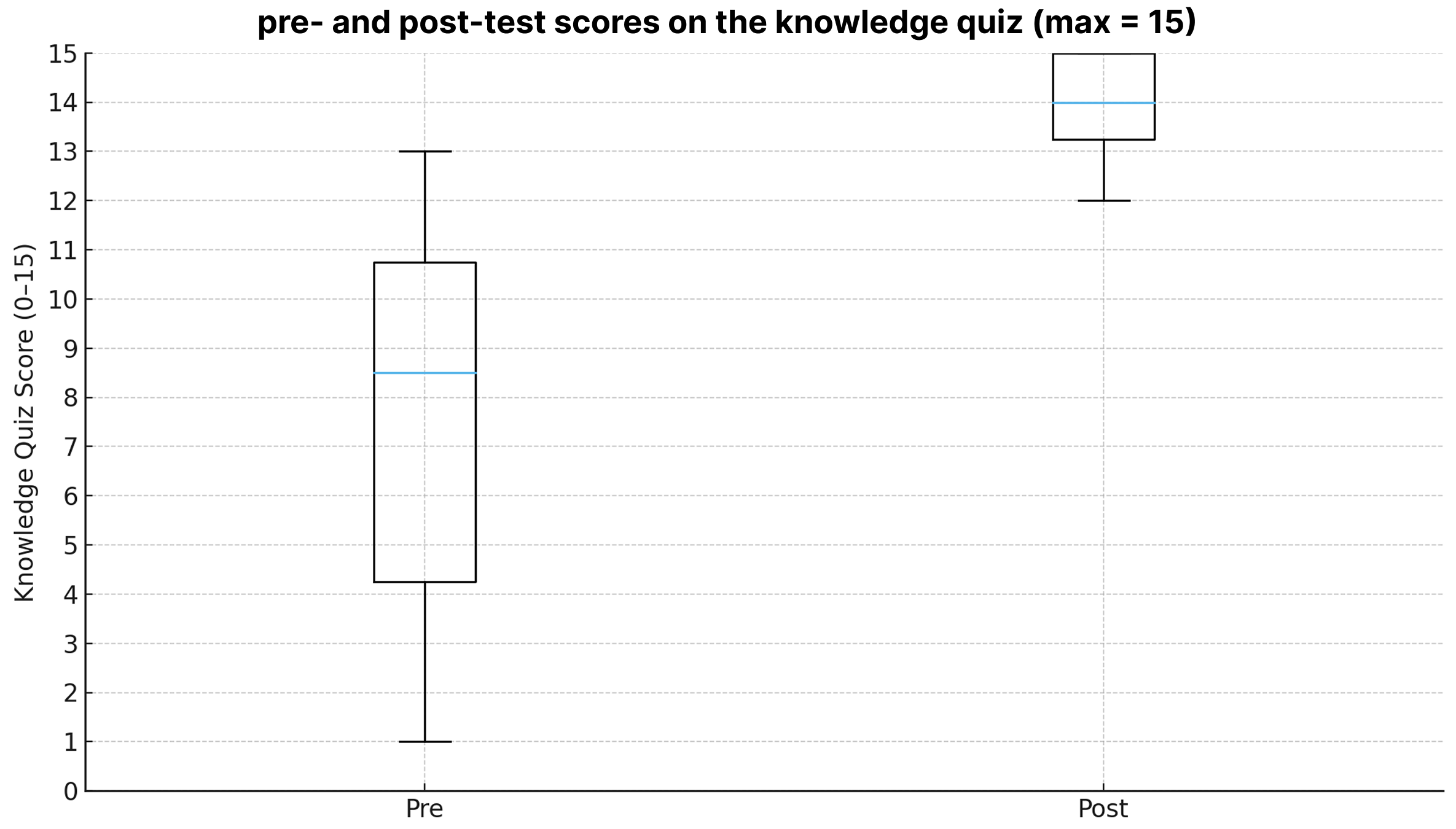}
 \caption{Distributions of pre- and post-test scores on the 15-item knowledge quiz.}
 \Description{Distributions of pre- and post-test scores on the 15-item knowledge quiz.}
    \label{prepost}
\end{figure}

After using the system, we assessed participants’ \textit{\textbf{immediate learning outcomes}} on Diaolou knowledge. All participants improved their scores, with an average gain of 6.11 items (see Figure~\ref{prepost}).
Participants’ knowledge quiz scores increased from $M_\text{pre} = 7.78$ ($SD = 3.81$) to $M_\text{post} = 13.89$ ($SD = 1.08$), yielding a mean gain of 6.11 items ($SD = 4.17$). A paired-samples $t$-test confirmed that this improvement was statistically significant, $t(17) = 6.22$, $p < .001$. However, six participants ($P1, P3, P5, P8, P12$, and $P13$) reached the maximum post-test score (15/15), which suggests a ceiling effect and limits the sensitivity of the quiz to additional learning gains among higher-performing participants.

We therefore interpret these results primarily as evidence of substantial factual learning rather than fine-grained differentiation between participants, and we refined the item design in Study~2 (see Section~\ref{S-2}) to increase difficulty and mitigate such ceiling effects.

\subsubsection{Impact on System Experience.}

\begin{figure*}[t]
  \centering
  \begin{subfigure}[t]{0.49\textwidth}
    \centering
    \includegraphics[width=\linewidth]{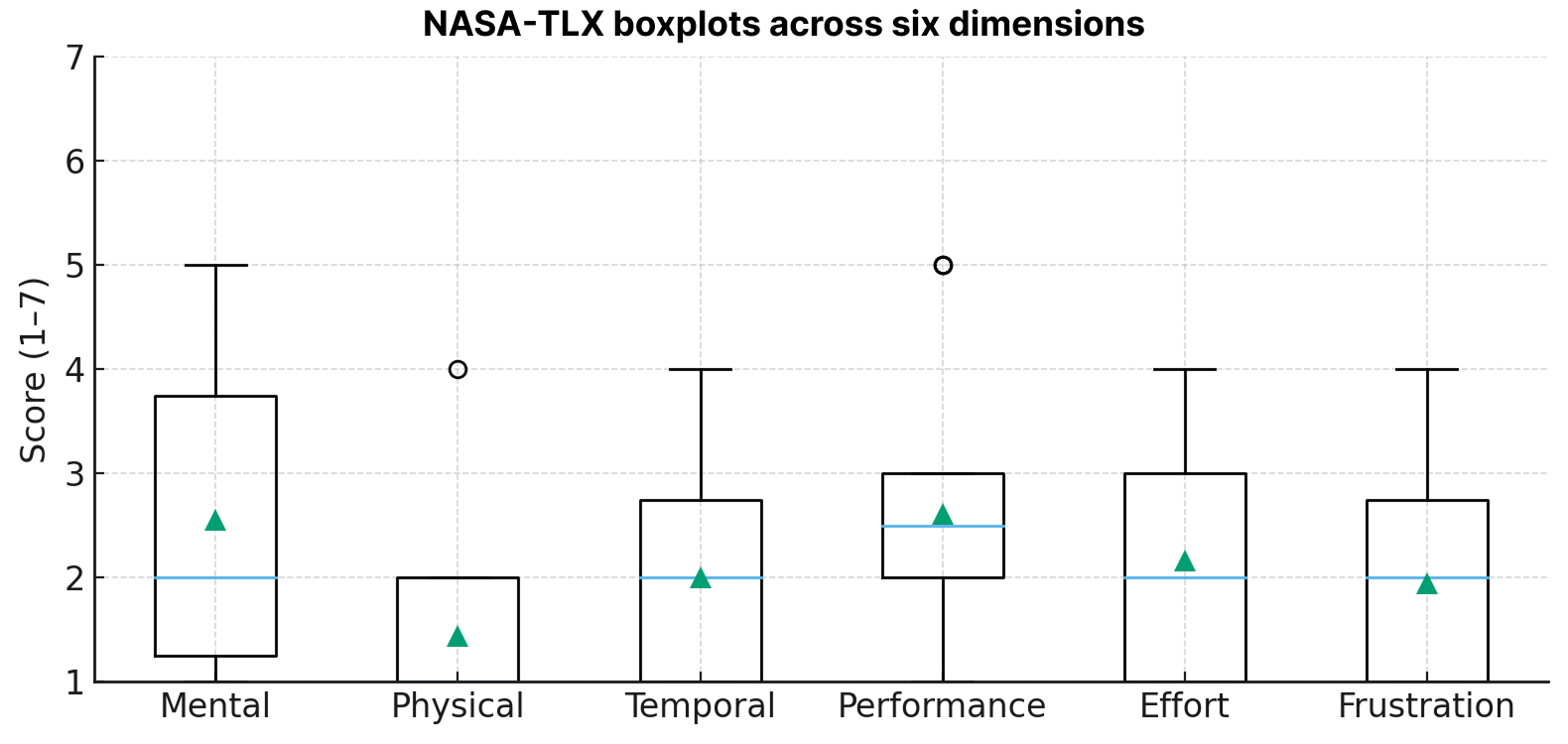}
    \caption{NASA-TLX ratings across six dimensions (1=best, 7=worst). The triangle indicates the mean.}
    \label{fig:nasa_results}
  \end{subfigure}
  \hfill
  \begin{subfigure}[t]{0.49\textwidth}
    \centering
    \includegraphics[width=\linewidth]{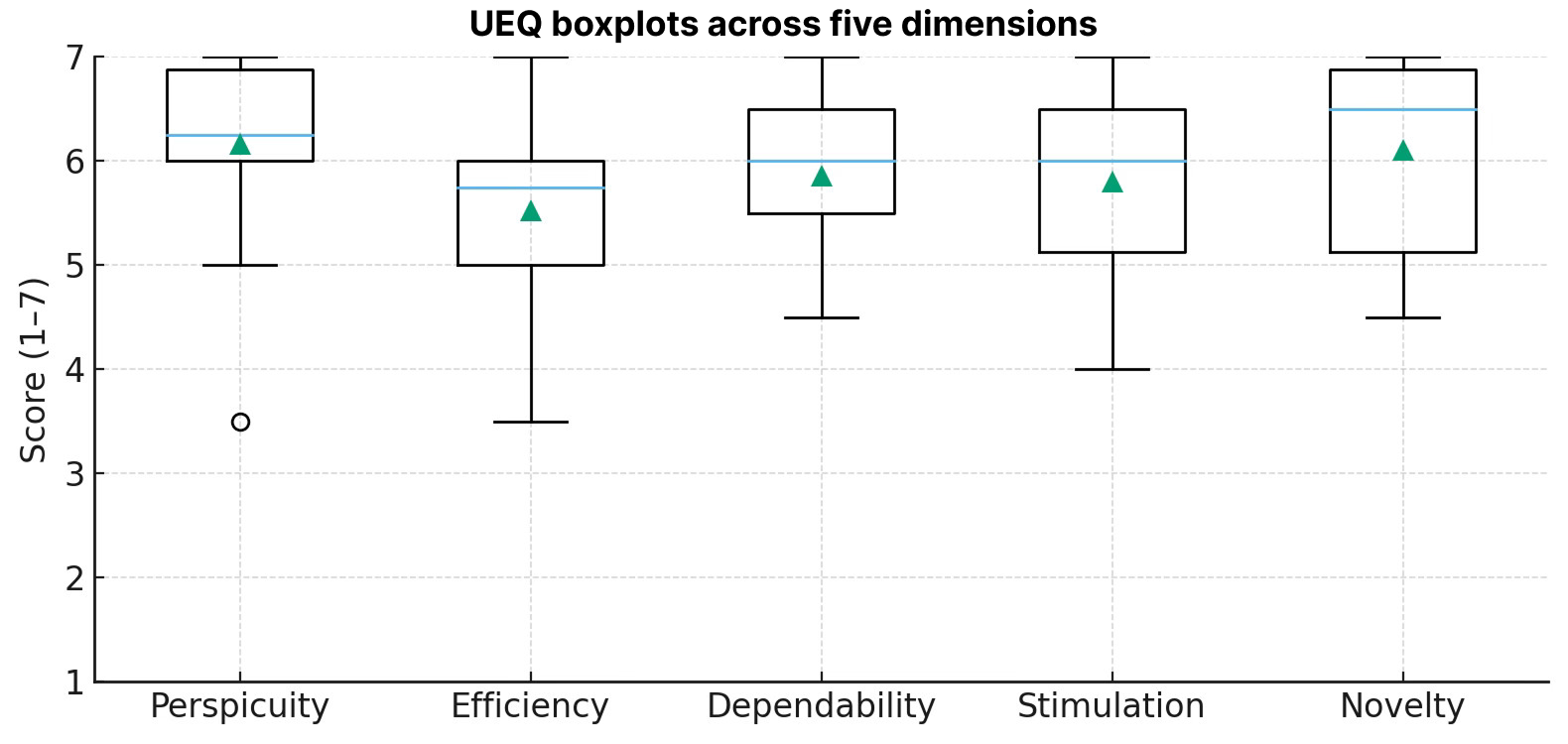}
    \caption{UEQ ratings across five dimensions (1=worst, 7=best). The triangle indicates the mean.}
    \label{fig:ueq_results}
  \end{subfigure}

  \caption{Workload and User Experience Ratings Across Conditions.}
  \Description{Two side-by-side plots: (a) NASA-TLX ratings across six dimensions (lower is better) with a mean marker; (b) UEQ ratings across five dimensions (higher is better) with a mean marker.}
  \label{fig:nasa_ueq}
\end{figure*}

\textit{\textbf{System usability.}}
We tested whether each UEQ dimension exceeded the neutral midpoint of 4 on a 7-point Likert scale (see Figure~\ref{fig:ueq_results}). 
All five dimensions were significantly higher than 4 (one-sample $t$-tests; $n = 18$):
\textit{Perspicuity} ($M = 6.17$, $SD = 0.87$), $t(17) = 10.51$, $p < .001$, $d = 2.48$;
\textit{Efficiency} ($M = 5.69$, $SD = 0.75$), $t(17) = 9.58$, $p < .001$, $d = 2.26$;
\textit{Dependability} ($M = 5.64$, $SD = 0.61$), $t(17) = 11.33$, $p < .001$, $d = 2.67$;
\textit{Stimulation} ($M = 5.81$, $SD = 0.79$), $t(17) = 9.72$, $p < .001$, $d = 2.29$;
and \textit{Novelty} ($M = 6.11$, $SD = 0.87$), $t(17) = 10.33$, $p < .001$, $d = 2.44$.
A repeated-measures ANOVA with dimension as a within-subject factor showed a main effect of dimension,
$F(4. 68) = 4.13$, $p = 0.005$, $\eta_p^2 = 0.20$.

Paired-samples $t$-tests indicated that \textit{Perspicuity} and \textit{Novelty} were rated higher than \textit{Efficiency}
(\textit{Perspicuity} $>$ \textit{Efficiency}: $t(17) = 2.72$, $p_{\text{adj}} = 0.029$; 
\textit{Novelty} $>$ \textit{Efficiency}: $t(17) = 2.29$, $p_{\text{adj}} = 0.035$);
all other pairwise comparisons were non-significant after Bonferroni correction ($p_{\text{adj}} > 0.05$). The results suggest that participants found the system easy to learn and innovative, while indicating room for improving responsiveness.

\textit{\textbf{Cognitive Load.}} 
We conducted one-sample $t$-tests against the scale midpoint (4 on a 1--7 scale) to assess whether participants experienced the system as cognitively or physically demanding ($N=18$). All six NASA-TLX dimensions were significantly below the  midpoint, indicating consistently low perceived workload.

Physical demand was minimal ($M = 1.44$, $SD = 0.78$), and temporal demand was also low ($M = 2.00$, $SD = 1.03$); both were significantly below the midpoint of 4 (physical: $t(17) = -13.83$, $p < .001$; temporal: $t(17) = -8.25$, $p < .001$). Mental demand ($M = 2.56$, $SD = 1.46$; $t(17) = -4.19$, $p < .001$) and self-rated performance (lower scores indicate better performance; $M = 2.61$, $SD = 1.33$; $t(17) = -4.42$, $p < .001$) were numerically higher than the other subscales (i.e., closer to the midpoint), but still significantly below neutral. Effort ($M = 2.17$, $SD = 1.04$; $t(17) = -7.46$, $p < .001$) and frustration ($M = 1.94$, $SD = 0.94$; $t(17) = -9.30$, $p < .001$) likewise remained low and were significantly below the midpoint, indicating light cognitive and emotional workload overall (see Table~\ref{fig:nasa_results}). 

A repeated-measures ANOVA on the six NASA--TLX dimensions revealed a significant main effect of dimension, $F(5, 85) = 4.29$, $p = .002$, $\eta_p^2 = .20$. Post-hoc pairwise comparisons using paired-samples $t$-tests indicated that no pairwise differences remained significant after Bonferroni correction (all $p_{\text{adj}} > .05$). 
Overall, participants experienced the system as requiring low cognitive, temporal, and physical effort, with minimal frustration, indicating that the interaction imposed only light workload.

\begin{figure*}[ht]
\centering
\includegraphics [width=\textwidth]{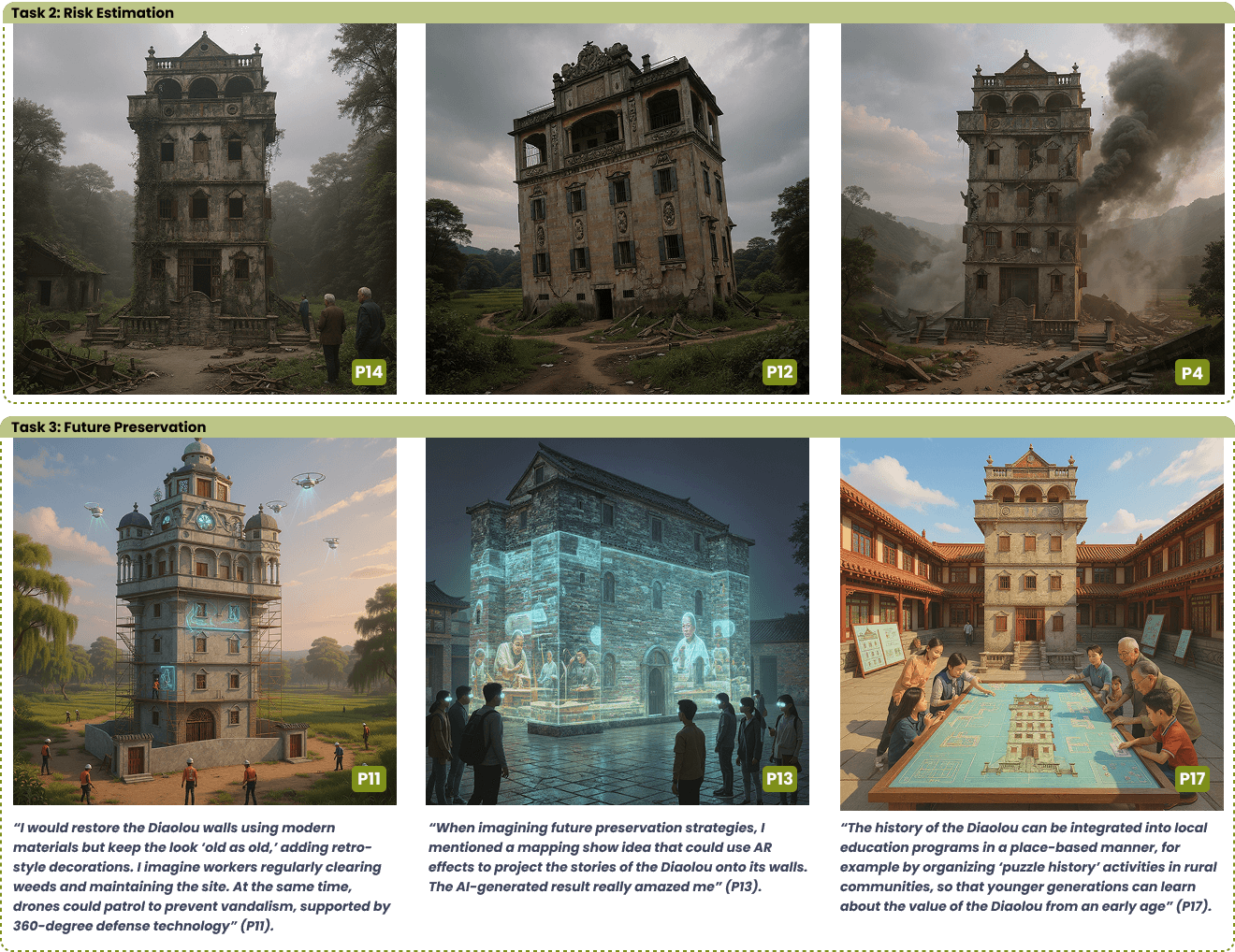} 
\caption{Generated images from the pilot study that participants deemed satisfactory: Task~2 \textit{Risk Estimation} scenes ($P14$, $P12$, $P4$) and Task~3 \textit{Future Preservation} scenes ($P11$, $P13$, $P17$), together with their reflections on the results.} 
\Description{Generated images from the pilot study that participants deemed satisfactory: Task 2 Risk Estimation scenes (P16, P12, P4) and Task 3 Future Preservation scenes (P11, P13, P17), together with their reflections on the results.}
\label{baohu} 
\end{figure*}

\subsubsection{Inquiry-Based Experience}
Several participants described the historical LLM-based conversational agent as making the experience more enjoyable and immersive, and their knowledge gains were accompanied by active, inquiry-based engagement with the system.

Several participants ($P2, P3, P5, P9, P10, P13$) actively posed questions to the agent to clarify uncertainties and explore historical details. For example, $P13$ queried the agent for additional alternatives and attempted to produce wall paintings that blended Chinese and Western elements, incorporating the Chinese character ``\textit{XI}'' (double happiness) and plant motifs to achieve the desired effect (see Figure~\ref{cover}(a)). As $P8$ described, 
\begin{quote}
\textit{``In the background introduction part, being guided by Huang Bixiu's
story helped me better understand the stories behind the Kaiping Diaolou through his lived
experience.''} 
\end{quote}

These findings suggest that the system not only improved factual recall but also encouraged participants to engage in open-ended exploration and meaning-making around Diaolou heritage.

\begin{figure*}[ht]
\centering
\includegraphics [width=\textwidth]{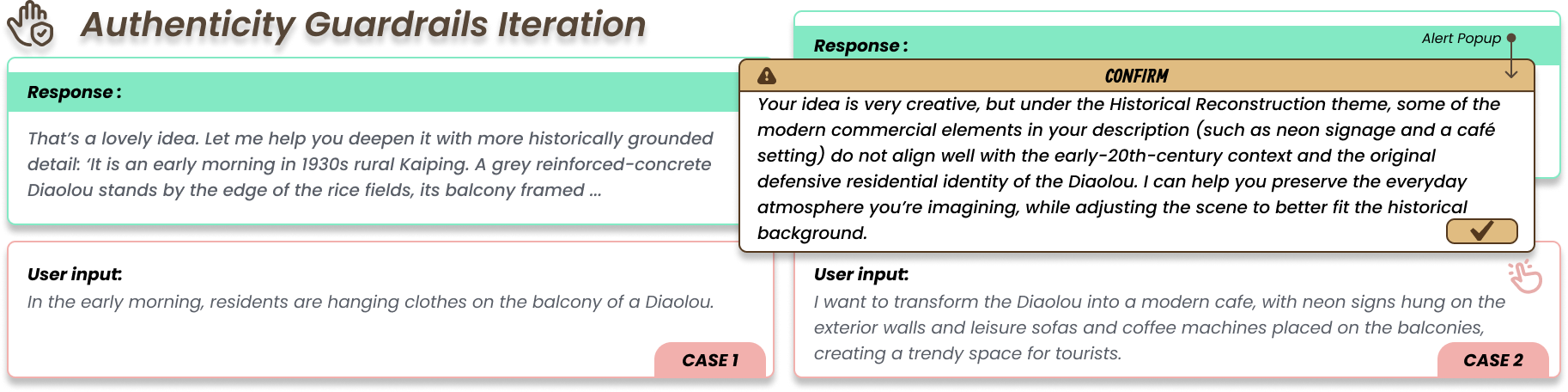} 
\caption{Examples of the iterative authenticity guardrails in action. The system deepens historically aligned ideas into narrative scene descriptions (Case 1), and flags anachronisms or incompatible cultural elements by displaying an popup with explanatory guidance and suggested alternatives (Case 2).}
\Description{Examples of the iterative Authenticity Guardrails in action. The system deepens historically aligned ideas into narrative scene descriptions (Case 1), and flags anachronisms or incompatible cultural elements by displaying an popup with explanatory guidance and suggested alternatives (Case 2).}
\label{pop} 
\end{figure*}

\subsubsection{Authenticity Guardrails Ensured Authentic Reconstruction.}\label{2.2}


Participants generally perceived the guardrails positively in interviews. For example, $P1$ noted,
\begin{quote}
\textit{“I could clearly see which parts were evidence-based and which were imaginative; these changes stopped me from making careless claims.”}
\end{quote}
At the same time, some participants (e.g., $P2$, $P4$, $P12$) also encountered situations where the system’s authenticity guardrails constrained their ideas. When prompts contained anachronistic or fantastical elements during \textit{Historical Reconstruction}, the system silently filtered or removed them; for instance, $P12$ reported that an \textit{“alien attack”} description in \textit{Task~1} was not executed. Similarly, $P4$ noted,
\begin{quote}
\textit{“I described a war scene with tanks and armored vehicles, but the prompt did not transform according to my intention.”}
\end{quote}

Most participants acknowledged that such guardrails protected \textit{historical credibility}, \textit{“Especially for themes that involve looking back to the past, I consider this mechanism (authenticity guardrails) effective for certain topics; without such constraints, it could create cultural conflicts in museum contexts”} ($P18$).

In another case, $P1$ and $P16$ recommended proactively flagging prompts that fall outside the historical setting and explaining the constraint: \textit{``Some users may not realize this is a constraint and might assume the AI malfunctioned''}($P16$). $P12$ added that the interface should provide clearer, task-specific guidance for each task to reduce ambiguity and better support user decision-making.

\subsubsection{Iterative Scaffolding Supported Both Alignment and Elaboration}

All participants successfully completed the required tasks within the allotted time. 
Cumulatively across the three tasks, participants averaged 4.7 prompt iterations and 18.9 generated images per person, with each iteration generating four images (see Table~\ref{tab:aigc_task} in Appendix~\ref{TaskPerformance}). 
Interviews did not suggest a clear relationship between the number of prompt iterations and perceived task or system difficulty; rather, iteration depth (i.e., number of prompt iterations) was idiosyncratic, reflecting individual creative strategies, desired fidelity, and exploration styles.
Representative task outputs are illustrated in Figure~\ref{cover}.

While many participants reached a satisfactory result in a single iteration, some chose to engage in multiple rounds of prompting to better describe timing, calibrate hazard intensity, and add scene details. 
Six participants ($P1, P3, P8, P13, P16, P18$) conducted two or more rounds to both align the generated output with their intent and elaborate on visual and narrative details. 
For example, $P3$ refined their prompt over three iterations to depict the urbanization risks faced by the Diaolou in the future, noting: 
\textit{“I felt that the prompt scaffolding could build on my previous ideas and gradually reach an effect I was satisfied with.”} 
As $P13$ added, 
\begin{quote} 
\textit{“Overall, I was satisfied that the prompt accurately captured my core ideas and generated them effectively; when I wanted to try different styles, it also provided timely feedback.”} 
\end{quote} 

\subsubsection{Simulated Risks Catalyzing Preservation Awareness and User-Driven Safeguarding Strategies}\label{3.2}

Participants leveraged the \textit{GenAI Module} to create risk scenarios (e.g., \textit{abandonment, structural collapse, flooding, urbanization, over-commercialization}) they perceived as realistic, and in doing so, they articulated a strong sense of concern for the Diaolou.
Participants felt that these visualizations made potential risks more \textit{tangible and emotionally resonant} (see Figure~\ref{baohu}, Task~1). 

For instance, $P4$ depicted \textit{“a Diaolou collapsing, with smoke and rubble swirling around it”}, while $P16$ and $P14$ imagined homecoming scenes in which elderly villagers revisit their former homes, foregrounding loss and memory. As $P14$ explained, \textit{“I imagined an old couple standing in front of their house, looking back on their lives with sadness.”}

Participants then put forward a series of action-oriented proposals for the safeguarding of the Diaolou, with a particular emphasis on the importance of \textit{community engagement} and the role of \textit{digital technologies} in the preservation of the cultural value of the site (see Figure~\ref{baohu}, Task~2).
For example, 
    
\begin{quote} 
\textit{“The history of the Diaolou can be integrated into local education programs in a place-based manner, for example by organizing ‘puzzle history’ activities in rural communities, so that younger generations can learn about the value of the Diaolou from an early age”} ($P17$). 
\end{quote}
Other participants highlighted the use of digital technologies: ($P11$) 
AR projection mapping ($P13$),
and XR-based digital reconstructions ($P16$). 

Notably, some participants ($P5$, $P11$, $P14$, $P18$) suggested integrating preservation cases from other World Heritage sites as references. Taken together, \textit{Gen-Diaolou} scaffolded a progression from \emph{risk awareness} to \emph{actionable preservation strategies}, linking visceral scenario-making with concrete, community- and technology-enabled interventions.

\subsection{Feedback and Iteration}\label{Iterative}

Building on the feedback from participants, we identified several iteration suggestions to further improve \textit{Gen-Diaolou}’s usability, creative support, and educational value (see Table~\ref{tab:iteration} in Appendix~\ref{FI}). These implications highlight concrete directions for refining the interaction flow, strengthening knowledge–creation integration, enhancing guardrail transparency (see Figure~\ref{pop}), and enriching multimodal immersion. We then used to derive actionable design improvements for subsequent field study.

\begin{figure*}[ht]
\centering
\includegraphics [width=\textwidth]{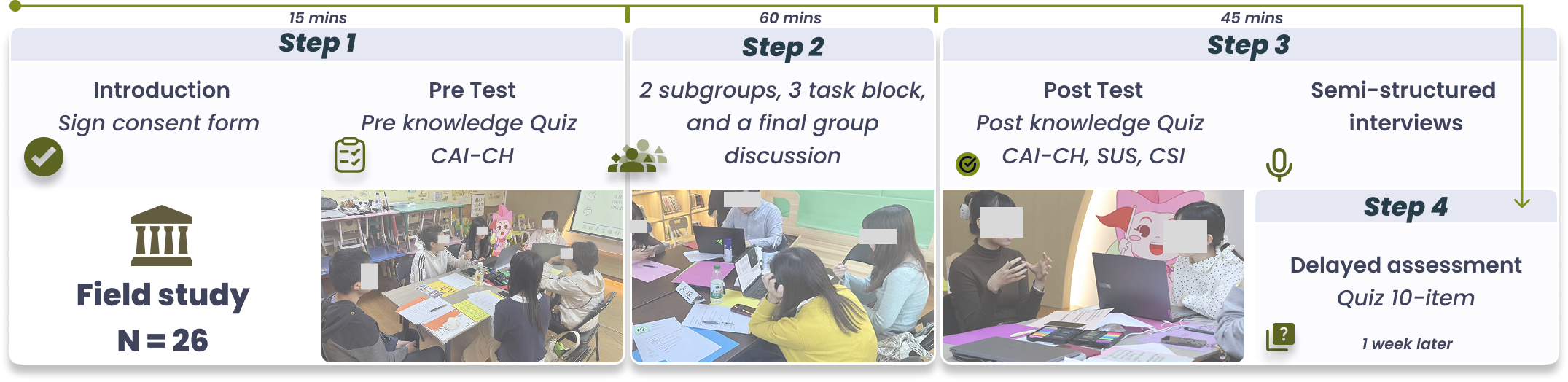} 
\caption{Field study procedure diagram with photos from the workshop. }
\Description{Field study procedure diagram with photos from the workshop. }
\label{fieldfig} 
\end{figure*}

\section{Study 2: Field Study}\label{S-2}

To enhance external validity in a real-world museum setting, we conducted a between-subjects field experiment at a local museum\footnote{Museum introduction for the \emph{Jiangmen Wuyi Museum of Overseas Chinese}: \url{https://www.prdculture.org.cn/ygawlzxwen/greater/202312/afc7dff327324cbcafbc805a3e4438eb.shtml}.} with a diverse participant pool ($N = 26$), including local community members and visiting tourists (refer to Section~\ref{survey}). This study assessed the system following iterative refinements (refer to Section~\ref{Iterative}) and aimed to isolate the incremental contribution of the \textit{GenAI Module}. We compared two conditions using the same \textit{Gen-Diaolou} interface and content: the \textbf{Base} condition and the \textbf{Learn+GenAI} condition. 

{\color{black}Museum management reviewed the study flow and on-site procedures to assess feasibility and visitor safety within the workshop setting. A museum staff member participated in the on-site sessions to support visitor outreach and logistics during the recruitment and workshop activities. Participants received souvenirs valued at approximately 100 CNY (about USD 14.3), along with a museum-provided CH-themed book, as compensation. }

\subsection{Participants and Procedure}\label{survey}

We recruited 26 participants (23 female, 3 male; aged 20–43 years, $M = 30.08$, $SD = 8.03$)\footnote{Further details on participant demographics are provided in Appendix~A} through social media and on-site posters at museum. All participants provided informed consent. Participants represented a diverse range of backgrounds: 12 were local residents, 10 were tourists from other provinces, and 4 were visitors from other cities within the same province. 

We collected baseline survey measures, including participants’ background, self-rated GenAI proficiency, a 20-item Diaolou knowledge quiz, and a custom 10-item \textit{Conservation Awareness Index for Cultural Heritage} (CAI-CH).

Based on these baseline measures, participants were allocated using stratified block randomisation into two experimental conditions: the \textbf{Base} condition (\textit{GenAI Module} disabled; $F1 - F13$) and the \textbf{Learn+GenAI} condition (with both the \textit{Knowledge Module} and the \textit{GenAI Module}; $F14 - F26$). 
Each condition was held in a separate session. 

The procedure comprised four phases (see Figure~\ref{fieldfig}): introduction, task block, post measures, and delayed assessment:

\textit{\textbf{Introduction (15\,min).}}
Participants were informed about the workshop background, system interface, task goals, task card distribution, and session timing. All participants (13 in each condition) were then divided into two subgroups, with each subgroup being supported by a research assistant who ensured consistent pacing and facilitated the task and discussions. Each group was provided with colored paper, markers, and laptops.

\textit{\textbf{Task block (60\,min).}}
Before starting the tasks, participants were given time to freely explore the system so that they could familiarize themselves with the interface and understand the available modules. After this initial exploration, participants proceeded to complete three themed task blocks (\textit{Historical Reconstruction}, \textit{Risk Estimation}, and \textit{Future Preservation}). 
For the \textbf{Base} condition, participants first used the \textit{Knowledge Module} to explore the content and then expressed their ideas for the three themed scenarios through sketches and brief written descriptions. For the \textbf{Learn+GenAI} condition, after completing the same \textit{Knowledge Module} step, participants then used the \textit{GenAI Module} to generate at least one satisfactory image for each task theme. After completing the creative tasks for each theme, all participants took part in a brief group discussion.

\textit{\textbf{Post measures (45\,min).}}
Participants completed a post Diaolou knowledge quiz and the SUS, CSI and CAI-CH. This was followed by a semi-structured interview conducted by two researchers (see Section~\ref{EvaluationDimensions}).

\textit{\textbf{Delayed assessment (1 week later).}}
One week after the workshop, participants received a link to a 10-item quiz (with optional transfer items) via the group chat. They completed the quiz online.

\subsection{Evaluation Dimensions}\label{EvaluationDimensions}

\subsubsection{Learning Outcomes}
Building on the pilot study (see Section~\ref{q}), we refined and expanded the custom Diaolou knowledge quiz in consultation with two domain experts. We assessed \textbf{immediate learning outcomes} using a 20-item multiple-choice test (10 \textbf{factual} and 10 \textbf{conceptual} questions), targeting recall and understanding, respectively, and administered before and after the visit. Knowledge retention was assessed with a 10-item multiple-choice \textbf{delayed assessment}, delivered as an online survey one week after the visit. All quiz items are provided in the supplementary material.

\subsubsection{Subjective Experience}
\textbf{Heritage conservation awareness} was measured pre- and post-visit using a custom 10-item, five-dimension \textit{Conservation Awareness Index for Cultural Heritage} (CAI-CH). Items were constructed with reference to UNESCO’s \textit{Operational Guidelines for the Implementation of the World Heritage Convention}\footnote{UNESCO The Operational Guidelines
for the Implementation of
the World Heritage Convention. Available at~\url{https://whc.unesco.org/en/guidelines/}.} and de la Torre et al.’s \textit{values and heritage conservation} framework~\cite{de2013values, ashworth2007pluralising, waterton2010recognition}, which emphasizes the protection, conservation, and transmission of cultural properties to future generationsm.
The \textit{System Usability Scale} (SUS)~\cite{sus} assesses overall \textbf{system usability}, whereas the \textit{Creativity Support Index} (CSI)~\cite{cherry2014quantifying} evaluates how well the system supports participants' \textbf{creative processes}. We adapted six SUS items to a 7-point scale and analyzed them dimension-wise rather than computing the canonical SUS score. All details are listed in supplementary materials.


\subsubsection{Qualitative Data}
We collected qualitative feedback via post-session group discussions and semi-structured interviews. All sessions were audio-recorded, transcribed, and anonymized. {\color{black}We conducted reflexive thematic analysis \cite{clarke2017thematic}: two authors co-coded an initial subset to calibrate a shared coding scheme, then iteratively coded the remaining transcripts and consolidated themes through regular discussions; disagreements were resolved through discussion with periodic peer debriefs.}

\subsection{Findings}

\begin{figure*}[t]
  \centering

  \begin{subfigure}[b]{0.25\textwidth}
    \centering
    \includegraphics[width=\linewidth]{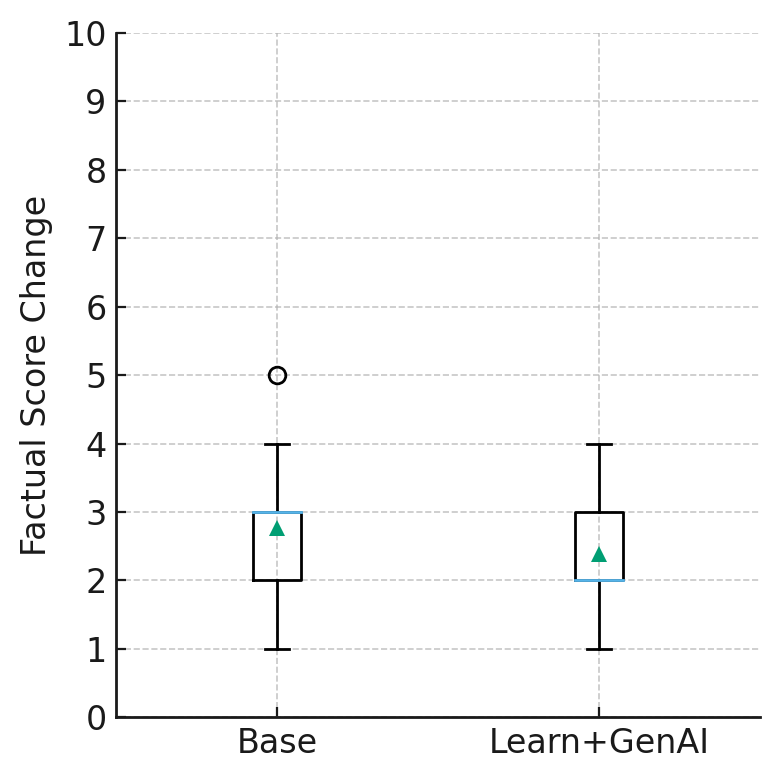}
    \caption{}
    \label{fig:conceptual-change}
  \end{subfigure}
  \hfill
  \begin{subfigure}[b]{0.25\textwidth}
    \centering
    \includegraphics[width=\linewidth]{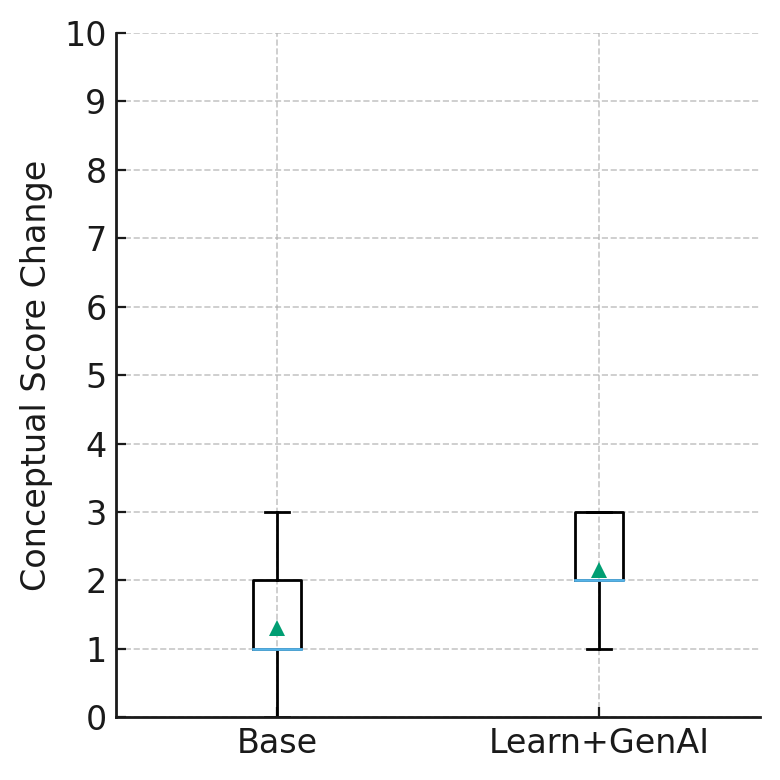}
    \caption{}
    \label{fig:factual-change}
  \end{subfigure}
  \hfill
  \begin{subfigure}[b]{0.25\textwidth}
    \centering
    \includegraphics[width=\linewidth]{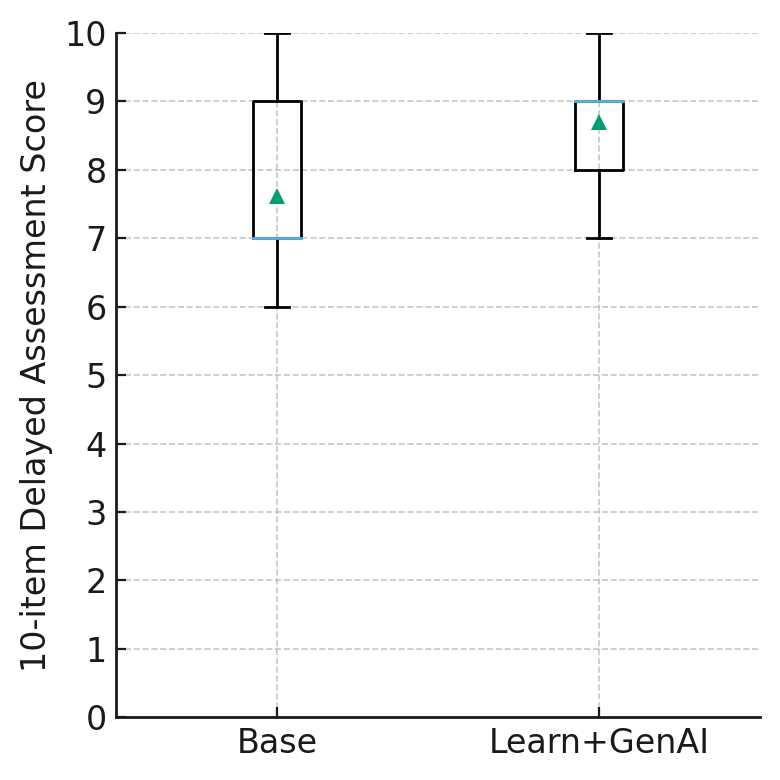}
    \caption{}
    \label{fig:delayed-assess}
  \end{subfigure}

  \caption{Participants’ learning outcomes across the Base and Learn+GenAI conditions:
  (a) change in factual knowledge scores, (b) change in conceptual knowledge scores, and (c) delayed post-test performance (all scores out of 10).}
  \Description{Three side-by-side plots showing participants’ learning outcomes: (a) change in factual knowledge scores, (b) change in conceptual knowledge scores, and (c) delayed post-test performance (all scores out of 10). The triangle denotes the mean.}
  \label{fig:conceptual-factual-change}
\end{figure*}

\subsubsection{Impact of Learning Outcomes}

To examine differences in learning outcomes across conditions, we analyzed scores from both the immediate learning assessment and the delayed assessment. The 20-item knowledge quiz demonstrated acceptable internal consistency (KR-20 = .83 at pre-test; .85 at post-test). The 10-item delayed assessment also showed acceptable reliability (KR-20 = .79), indicating that the items consistently measured participants’ Diaolou-related knowledge across administrations.

\textit{\textbf{Immediate learning outcomes assessment.}}
We first analyzed pre–post changes in the knowledge quiz across conditions. For \textbf{factual knowledge}, both conditions showed clear learning gains. Participants in the \textbf{Base} condition improved by an average of $2.77$ items ($SD = 1.09$), while those in the \textbf{Learn+GenAI} condition improved by $2.38$ items ($SD = 1.04$) (see Figure~\ref{fig:conceptual-change}). An independent-samples Welch's $t$-test indicated no statistically significant difference between conditions on factual improvements, $t(23.95) = 0.92$, $p = .368$, $d = 0.37$. This suggests that the two conditions were similarly effective for supporting recall of Diaolou-related factual information. 

For \textbf{conceptual knowledge}, both conditions again showed positive gains, but the \textbf{Learn+GenAI} condition provided a clear advantage  (see Figure~\ref{fig:factual-change}). In the \textbf{Base} condition, conceptual scores increased by $M = 1.31$ items ($SD = 0.75$), whereas in the \textbf{Learn+GenAI} condition the mean gain was $M = 2.15$ items ($SD = 0.80$). The difference between conditions was statistically significant and large in magnitude, $t(23.88) = -2.78$, $p = .010$, Cohen's $d = 1.08$. Thus, while basic factual learning was comparable across conditions, the interactive setting with the \textit{GenAI Module} substantially improved participants' interpretive understanding of Diaolou, aligning with our goal of supporting higher-level, interpretation-oriented heritage learning.

\textit{\textbf{Delayed assessment.}}
A week after the visit, both groups demonstrated good delayed retention on a 10-item knowledge assessment, with relatively high mean scores in both conditions (see Figure~\ref{fig:conceptual-factual-change}(c)). The \textbf{Learn+GenAI} condition ($M = 8.77$, $SD = 0.93$) scored higher than the \textbf{Base} condition ($M = 7.46$, $SD = 1.05$). An independent-samples Welch's $t$-test showed that this difference was statistically significant, $t(23.64) = 3.37$, $p = .003$, with a large effect size (Cohen’s $d = 1.32$), indicating better knowledge retention for participants in the \textbf{Learn+GenAI} condition.

\begin{table*}[t]
\centering
\caption{Pre--post changes in CAI-CH dimensions by condition (Base vs.\ Learn+GenAI).}
\Description{Pre--post changes in CAI-CH dimensions by condition (Base vs.\ Learn+GenAI), reporting mean (standard deviation), p-values, and significance codes.}
\label{tab:cai_prepost}

\setlength{\tabcolsep}{4pt}
\renewcommand{\arraystretch}{1.05}

\begin{tabular}{@{}%
  >{\raggedright\arraybackslash}p{5.0cm}
  l
  >{\centering\arraybackslash}p{2.3cm}@{\hspace{2pt}}
  >{\centering\arraybackslash}p{2.3cm}
  >{\centering\arraybackslash}p{1.6cm}
  c@{}}
\toprule
\textbf{Dimension} & \textbf{Condition} & \textbf{Pre $M(SD)$} & \textbf{Post $M(SD)$} & \textbf{$p$-value} & \textbf{Sig.} \\
\midrule
\multirow{2}{*}{\makecell[l]{\textbf{Value Recognition of}\\\textbf{Cultural Heritage}}}
  & Base        & 5.23 (0.70) & 5.54 (0.80) & .071      & n.s. \\
  & Learn+GenAI & 5.38 (0.68) & 6.15 (0.75) & .002      & **   \\
\midrule
\multirow{2}{*}{\makecell[l]{\textbf{Heritage Identity and}\\\textbf{Sense of Belonging}}}
  & Base        & 5.04 (0.85) & 5.46 (0.56) & .144      & n.s. \\
  & Learn+GenAI & 4.73 (0.63) & 6.15 (0.80) & $<.001$   & ***  \\
\midrule
\multirow{2}{*}{\makecell[l]{\textbf{Awareness of Intangible Heritage}\\\textbf{Living Traditions}}}
  & Base        & 4.69 (0.83) & 5.27 (0.67) & .054      & n.s. \\
  & Learn+GenAI & 4.85 (0.97) & 5.96 (0.56) & .005      & **   \\
\midrule
\multirow{2}{*}{\makecell[l]{\textbf{Willingness to Participate}\\\textbf{Public Engagement}}}
  & Base        & 4.77 (0.73) & 5.00 (0.71) & .337      & n.s. \\
  & Learn+GenAI & 5.04 (0.83) & 6.12 (0.68) & .006      & **   \\
\midrule
\multirow{2}{*}{\makecell[l]{\textbf{Cultural Sustainability and}\\\textbf{Intergenerational Responsibility}}}
  & Base        & 5.23 (0.86) & 5.65 (0.55) & .128      & n.s. \\
  & Learn+GenAI & 5.42 (0.57) & 6.23 (0.70) & .004      & **   \\
\bottomrule
\end{tabular}

\caption*{\small \textit{Note.} Significance codes: n.s.~$p \ge .050$, * $p<.050$, ** $p<.010$, *** $p<.001$.}
\end{table*}

\subsubsection{Impact of Preservation Awareness}
The \textit{Gen-Diaolou} interactive experience led to increased \textbf{CH preservation awareness} in both conditions, as reflected in higher post-visit mean scores across the five CAI-CH dimensions for both the \textbf{Base} and the \textbf{Learn+GenAI} conditions (see Table~\ref{tab:cai_prepost}). 
Across all five CAI-CH dimensions, both conditions showed pre–post improvements, but gains were consistently larger in the \textbf{Learn+GenAI} condition. 

For each condition, we conducted paired-samples $t$-tests comparing pre- and post scores on the five dimensions of the CAI-CH. 
In the \textbf{Base} condition, none of the pre--post differences reached statistical significance, although we observed a marginal increase in \textit{Awareness of Intangible Heritage and Living Traditions}, $t(12) = 2.14$, $p=.054$, while the remaining dimensions showed smaller, non-significant gains (all $p\geq .071$). 

In contrast, the \textbf{Learn+GenAI} condition exhibited significant improvements across all five dimensions: \textit{Value Recognition of Cultural Heritage}, $t(12) = 3.93$, $p=.002$; \textit{Heritage Identity and Sense of Belonging}, $t(12) = 4.32$, $p<.001$; \textit{Awareness of Intangible Heritage and Living Traditions}, $t(12) = 3.43$, $p=.005$; \textit{Willingness to Participate and Public Engagement}, $t(12) = 3.33$, $p=.006$; and \textit{Cultural Sustainability and Intergenerational Responsibility}, $t(12) = 3.55$, $p=.004$. Taken together, these results suggest that the \textbf{Learn+GenAI} condition more effectively fostered visitors' preservation awareness and related attitudes than the Base condition.

\subsubsection{Impact on System Experience}\label{dingliang}

The perception of \textit{Gen-Diaolou} was assessed using the SUS and CSI scales, and scores were compared between two conditions (see Table~\ref{tab:user_feedback}). 
To investigate differences between groups in more detail, we conducted independent-samples Welch's $t$-tests on each SUS and CSI dimension and additionally computed effect sizes (Cohen’s $d$) to quantify the magnitude of these differences.

\begin{table*}[t]
\centering
\caption{Statistical user measures comparing the \emph{Base} and \emph{Learn+GenAI} conditions.}
\Description{Statistical comparison of user measures between the Base and Learn+GenAI conditions, including SUS, CSI, knowledge quiz improvement, and delayed knowledge assessment (means, standard deviations, and p-values with significance levels).}
\label{tab:user_feedback}
\setlength{\tabcolsep}{6pt}  
\renewcommand{\arraystretch}{1.08}

\begin{tabular}{@{}%
  >{\raggedright\arraybackslash}p{3.6cm}
  >{\raggedright\arraybackslash}p{3.0cm}
  >{\centering\arraybackslash}p{2.5cm}
  >{\centering\arraybackslash}p{3.0cm}
  >{\centering\arraybackslash}p{1.5cm}
  c@{}}
\toprule
\textbf{Categories} & \textbf{Factors} & \textbf{Base $M(SD)$} & \textbf{Learn+GenAI $M(SD)$} & \textbf{$p$-value} & \textbf{Sig.} \\
\midrule
\multirow{6}{*}{\makecell[l]{\textbf{System Usability Scale}\\\textbf{(SUS)}~\cite{sus}}}
  & Easy to use      & 5.54 (1.20) & 5.92 (0.64) & .321  & n.s. \\
  & Functions        & 4.62 (1.12) & 5.85 (0.80) & .005  & **   \\
  & Quick to learn   & 5.08 (1.38) & 6.08 (0.95) & .031  & *    \\
  & Learning curve   & 3.62 (0.96) & 1.85 (0.90) & .001  & ***  \\
  & Frequency        & 4.15 (1.14) & 5.92 (0.95) & .001  & ***  \\
  & Confidence       & 4.62 (1.39) & 6.15 (0.80) & .001  & ***  \\
\midrule
\multirow{6}{*}{\makecell[l]{\textbf{Creativity Support Index}\\\textbf{(CSI)}~\cite{cherry2014quantifying}}}
  & Enjoyment            & 5.06 (1.21) & 5.67 (0.96) & .074  & n.s. \\
  & Exploration          & 4.88 (1.15) & 5.90 (0.93) & .001  & ***  \\
  & Results Worth Effort & 4.29 (1.05) & 6.04 (1.15) & .001  & ***  \\
  & Expressiveness       & 3.97 (1.24) & 6.02 (0.99) & .001  & ***  \\
  & Collaboration        & 4.05 (1.59) & 5.15 (1.09) & .001  & ***  \\
  & Immersion            & 3.88 (1.29) & 5.98 (0.94) & .001  & ***  \\
\midrule
\multirow{2}{*}{\makecell[l]{\textbf{Knowledge Quiz}\\\textbf{CImprovement}}}
  & Factual questions    & 2.77 (1.09) & 2.38 (1.04) & .316  & n.s. \\
  & Conceptual questions & 1.31 (0.75) & 2.15 (0.80) & .010  & **   \\
\midrule
\textbf{Delayed assessment}
  & Knowledge retention  & 7.46 (1.05) & 8.77 (0.93) & .003  & **   \\
\bottomrule
\end{tabular}

\caption*{\small \textit{Note.} Significance codes: n.s.~$p \ge .050$, * $p<.050$, ** $p<.010$, *** $p<.001$.}
\end{table*}

\textit{\textbf{System usability.}} 
Participants rated both conditions as relatively \textit{easy to use} (all scores $> 4$ on a 7-point scale), with the \textbf{Learn+GenAI} condition rated slightly easier to use ($M = 5.92$, $SD = 0.64$) than the \textbf{Base} condition ($M = 5.54$, $SD = 1.20$), but this difference was not statistically significant 
($t(18.34) = 1.02$, $p = .321$, $d = 0.40$). In contrast, \textit{functionality} in the \textbf{Learn+GenAI} condition was evaluated as more comprehensive and better integrated ($M = 5.85$, $SD = 0.80$) than in the \textbf{Base} condition ($M = 4.62$, $SD = 1.12$), showing a statistically significant difference 
($t(21.72) = 3.22$, $p = .004$, $d = 1.26$).
\textit{Learnability} was also rated higher for the \textbf{Learn+GenAI} condition ($M = 6.08$, $SD = 0.95$) compared to the \textbf{Base} condition ($M = 5.08$, $SD = 1.38$), and this difference reached statistical significance 
($t(21.32) = 2.15$, $p = .043$, $d = 0.84$).
Additionally, the \textbf{Learn+GenAI} condition was perceived to have a substantially less demanding \textit{learning curve} ($M = 1.85$, $SD = 0.90$) than the \textbf{Base} condition ($M = 3.62$, $SD = 0.96$), indicating easier adoption and greater user-friendliness 
($t(23.89) = 4.85$, $p < .001$, $d = - 1.90$).
Participants further reported that they would use the \textbf{Learn+GenAI} system more \textit{frequently} ($M = 5.92$, $SD = 0.95$) than the \textbf{Base} system ($M = 4.15$, $SD = 1.14$); $t(23.25) = 4.28$, $p < .001$, $d = 1.68$) and felt greater \textit{confidence} while using it ($M = 6.15$, $SD = 0.80$ vs.\ $M = 4.62$, $SD = 1.39$); $t(19.20) = 3.46$, $p = .003$, $d = 1.36$).

\textit{\textbf{Creativity Support.}} 
Participants assigned to the \textbf{Learn+GenAI} condition tended to report 
higher CSI scores than those in the \textbf{Base} condition across all six 
dimensions (see table~\ref{tab:user_feedback}). A between-condition Welch independent-samples $t$-test indicated 
no statistically significant difference between conditions on 
\textit{enjoyment} ($M = 5.67$, $SD = 0.96$ vs.\ $M = 5.06$, $SD = 1.21$; $t(23.77) = 1.87$, $p = .074$, $d = 0.73$), suggesting that both conditions experienced the activity as generally fun and engaging. By contrast, the other five dimensions showed substantial between-condition differences favouring the \textbf{Learn+GenAI} condition: \textit{exploration}, 
$t(21.32) = 3.38$, $p < .001$, $d = 1.33$; \textit{results worth effort}, 
$t(22.93) = 5.99$, $p < .001$, $d = 2.35$; \textit{collaboration}, 
$t(19.18) = 3.78$, $p < .001$, $d = 1.48$; \textit{expressiveness}, 
$t(17.98) = 6.96$, $p < .001$, $d = 1.03$; and \textit{immersion}, 
$t(16.12) = 6.87$, $p < .001$, $d = 1.40$. These effects indicate that the 
GenAI-augmented system better supported trying out alternative ideas, 
producing outcomes that felt worth the effort, coordinating with collaborators, articulating ideas clearly, and feeling immersed in the activity. Overall, the \textit{Gen-Diaolou} \textbf{Learn+GenAI} condition demonstrated markedly stronger creativity support than the \textbf{Base} condition across the evaluated CSI dimensions.

\begin{figure*}[ht]
\centering
\includegraphics [width=\textwidth]{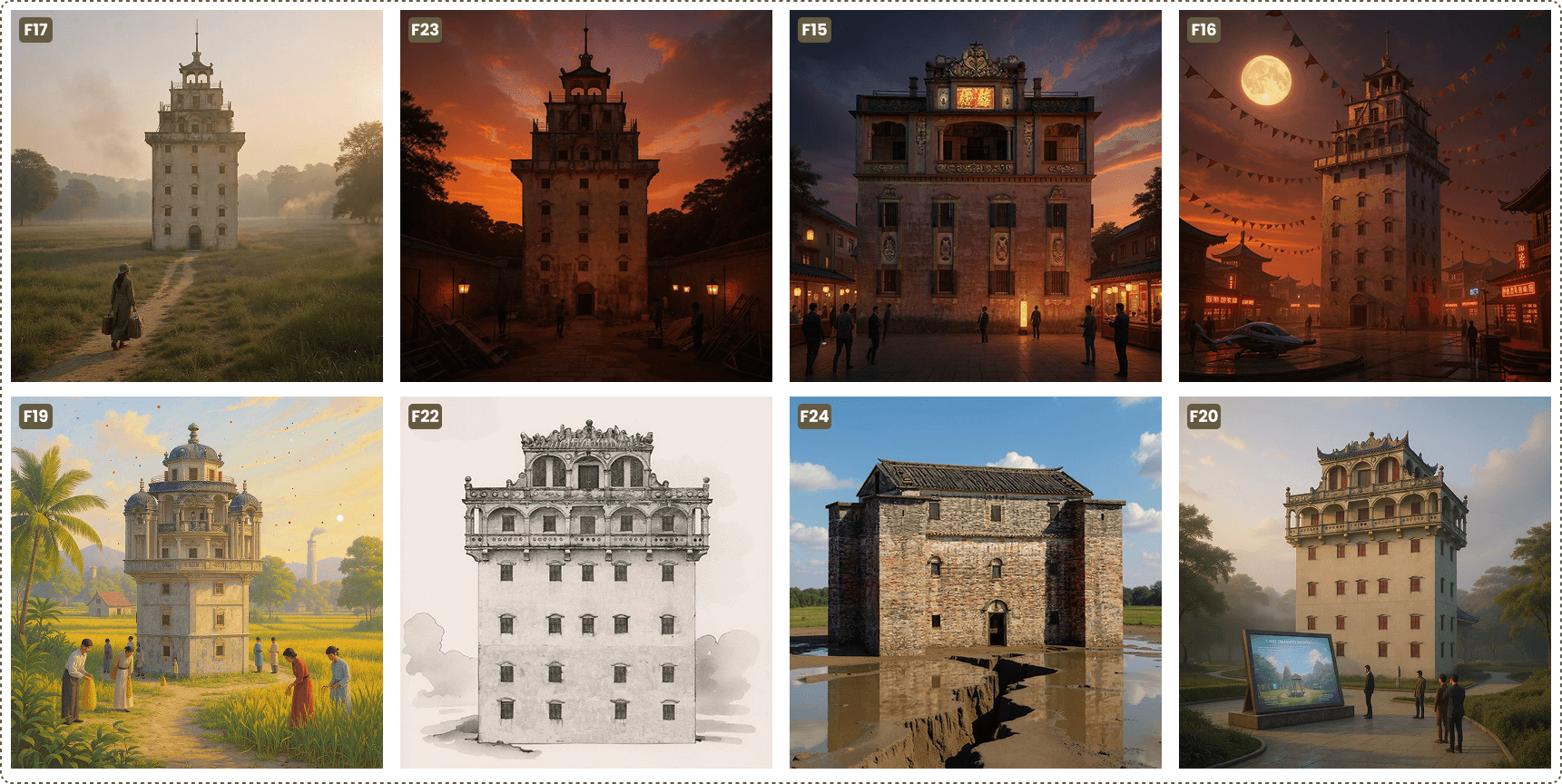} 
\caption{Generated images from the \textit{Learn+GenAI} condition deemed satisfactory by participants, including Task~1 \textit{Historical Reconstruction} ($F17, F19, F22, F23$), Task~2 \textit{Risk Estimationg} ($F15, F24$), and Task~3 \textit{Future Preservation} ($F16, F20$).} 
\Description{Generated images from the Learn+GenAI condition deemed satisfactory by participants, including Task 1 Historical Reconstruction (F17, F19, F22, F23), Task 2 Risk Estimation (F15, F24), and Task 3 Future Preservation (F16, F20).}
\label{FSfigure} 
\end{figure*}

\subsubsection{Qualitative Findings}

Building on our thematic analysis of the interviews, we organized participants’ interview data into four dimensions in this section: \textit{Knowledge Acquisition}, \textit{Creativity and Task Completion}, \textit{Preservation Awareness}, and \textit{Usability and Improvements}.

\textit{\textbf{Knowledge Acquisition.}}
Across both conditions, participants commonly noted that \textit{Gen-Diaolou} helped them gain a deeper appreciation of the Diaolou’s historical significance and architectural features. Several attributed this to the system’s \textit{knowledge structure} ($F1, F11, F23, F24$), \textit{conversational agent dialogue} ($F4, F9, F13, F21$) and \textit{prompt expansion} ($F14$, $F20$, $F25$ in the Learn+GenAI group), which they felt supported connections between past, present, and future perspectives.

\textit{Agent-guided dialogue enhanced knowledge integration.}
Several participants highlighted that the LLM-based conversational agent encouraged them to approach the Diaolou from historically appropriate viewpoints, shifting their attention from external aesthetics to embedded cultural meanings. For instance, 
$F13$ remarked:
\begin{quote}
\textit{“Talking with him (conversational agent) made the historical story feel much more vivid. Even as a local, I actually didn’t know who owned Ruishi Lou before, and this experience helped me understand the Diaolou more deeply.”} 
\end{quote}

\textit{Prompt expansion as subtle learning support.}
Some participants ($F14$, $F20$, $F25$) noted that the \textit{GenAI Module}’s prompt expansion (see Figure~\ref{pop}) supported learning by adding architectural details embedded in their ideas. As $F14$ put it:
\begin{quote} 
\textit{“When it turned my idea into a detailed description and then generated the image, I understood the past scene more clearly.”} \end{quote}
$F25$ added: \textit{“It felt like a kind of review for me, more or less.}

\textit{\textbf{Creativity and Task Completion.}}
Participants in the \textbf{Learn+GenAI} condition provided feedback on the GenAI features. Overall, they expressed satisfaction with the three tasks, although with varying levels of enthusiasm for different aspects. Most participants (10 out of 13) reported the highest satisfaction with the \textit{historical reconstruction}.

Participants generally responded positively to the way the system expanded their initial creative ideas.
Several participants reflected on moments when their ideas were flagged by the authenticity guardrails. 
\begin{quote}
\textit{“I imagined the Diaolou in an empty desert, and the system told me that didn’t match its historical village setting, then gave me a version with fields, houses, and ancestral halls.”} ($F21$).
\end{quote}
Most agreed that the mechanism helped preserve historical accuracy, while a few felt it sometimes made the process less flexible. As $F19$ noted, 
\begin{quote}
\textit{“If it triggers too often, I might lose patience.”} $F24$ suggested adding a \textit{mode with greater freedom}. 
\end{quote}

In Task 3 (\textit{Future Preservation}), participants proposed diverse future-oriented safeguarding strategies, such as using festival celebrations ($F16$) and interactive large-screen installations at the site to support engagement with the Diaolou ($F20$); illustrated in Figure~\ref{FSfigure}. As $F17$ remarked: \begin{quote}\textit{"I was really excited to see my idea materialized—the scene of people using drones to scan the Diaolou and measure the land."}\end{quote}

\textit{\textbf{Preservation Awareness.}}
A notable contrast emerged between conditions regarding heritage preservation awareness. In the \textbf{Base} condition, many participants reported difficulty emotionally connecting with potential risks or imagining themselves in roles that contribute to safeguarding the Diaolou. As $F1$ reflected, \begin{quote}\textit{“I mostly understood Diaolou protection as architectural restoration and policy work”}. \end{quote}

$F6$ and $F8$ suggested that heritage preservation should prioritize sustaining cultural memory and historical narratives rather than focusing solely on maintaining the physical structure. $F6$ noted:
\begin{quote}
\textit{“Honestly, I care more about the history and culture than the building itself. If our own children forget these memories one day, that’s the real tragedy. As for the structure, I feel it is good enough as long as it still stands."}
\end{quote}

By comparison, most participants in the \textbf{Learn+GenAI} condition (11 out of 13) felt that the Task~2 \textit{Risk Estimation} activities heightened their sensitivity to risks by helping them visualize structural decay, environmental threats, or inappropriate renovation scenarios. 

However, perspectives were nuanced. For instance, $F22$ remarked that the transformation of the site into a local cultural hotel \textit{“might not be a bad outcome,”} suggesting that some forms of adaptive reuse may also be perceived as meaningful continuity rather than cultural loss.

\textit{\textbf{Usability \& Improvements.}}
Participants from both conditions rated the usability of \textit{Gen-Diaolou} positively and also provided various suggestions for improving the system's usability and interactive features.
In the \textbf{Base} condition, participants highlighted the need for better guidance and interaction during the task phase. Most participants found it difficult to express their ideas by traditional means. For example, $F7$ and $F11$ mentioned the limitations of traditional methods. $F7$ noted:
\begin{quote}  
\textit{“To be honest, I didn’t really have a clear concept in mind. I felt I needed some guidance, because I wasn’t sure what I could actually do for the Diaolou.”}
\end{quote} 

In the \textbf{Learn+GenAI} condition,
participants (e.g., $F15$, $F16$) proposed that the system could better showcase user-generated content by allowing for more visibility on the homepage, through user voting or comments, encouraging user engagement and feedback. Additionally, the incorporation of a reward system ($F15$) was suggested to provide motivation throughout the creative process.

$F14$, $F18$, $F21$, and $F24$ also suggested adding voice interaction for inputting ideas, as they believed typing might be inconvenient for older users. 

Some participants expressed suggestions regarding the system's collaboration features, proposing \textit{multi-user collaboration} ($F18$) and \textit{cross-regional collaboration} ($F16$, $F23$). 

Furthermore, $F15$, $F16$ and $F26$ also suggested that the \textit{GenAI Module} should have more social features, as $F16$ remarked:
\begin{quote} 
\textit{"The content we created with AI should be displayed on the homepage, allowing people to like and comment. In some themes, many of our ideas might be similar, but today I noticed that others have proposed unique ways of expression, which should be saved and showcased"}.
\end{quote} 

\section{Discussion}

\subsection{Fostering Understanding of Cultural Heritage through AI-assisted Co-Creation}
Drawing on our formative study and two empirical studies, we conceptualize digital heritage engagement as a coupled cycle of participatory, learning, creating, and imagining. Learning provides conceptual grounding, creating turns knowledge into interpretation, and imagining future scenarios fosters agency and sustained engagement, positioning users as co-creators of evolving heritage meanings rather than passive receivers.

Previous studies have highlighted the value of using creative production to facilitate learning and understanding, proposing the “create-to-learn” paradigm~\cite{10.1145/3544548.3580999,createtolearn}. From an activity-theoretic perspective~\cite{kuutti1996activity}, this paradigm foregrounds learners’ active participation and shifts the focus from mere knowledge acquisition to engaged, meaning-making through creation. Design goals from formative work led to the layered architecture of the \textit{Knowledge Module} and \textit{GenAI Module}, which together position visitors not only as recipients of curated content but as \textbf{active interpreters} of CH. 

Results from the pilot study (Section~\ref{S-1}) showed that participants could translate guided learning into historically grounded creative visual outputs, with significant gains in knowledge recall. Participants used the system to experiment with different historical scenes and speculative futures, and interviews suggested that they were not only “getting the right answers” but also engaging in open-ended meaning-making around the Diaolou. 

In the field study, while the \textbf{Base} and \textbf{Learn+GenAI} conditions yielded comparable results in factual knowledge recall, the GenAI-augmented approach demonstrated significant advantages in conceptual understanding and knowledge retention.

\subsection{Promoting Cultural Heritage Preservation Awareness}

Our study suggests that \textit{Gen-Diaolou} links learning tasks with risk visualisation and future-oriented safeguarding ideation. Participants used the \textit{GenAI Module} to depict plausible risk scenarios (e.g., abandonment, structural collapse, flooding, over-commercialisation) and reported that these visualisations made threats more \textit{tangible and emotionally resonant}. Building on these scenes, they proposed community- and technology-driven safeguarding strategies (e.g., public education, participatory programmes, interactive media), indicating a shift from passive appreciation toward a more active sense of responsibility. 

This pattern aligns with our CAI-CH results, where the Learn+GenAI condition yielded stronger gains in value recognition, willingness to participate, and perceived responsibility than the \textbf{Base} condition, pointing to opportunities for GenAI systems that explicitly scaffold preservation-oriented reflection rather than focusing solely on engagement or creativity.

This aligns with prior research demonstrating that GenAI experience significantly augments both behavioral engagement and reflective processing compared to traditional non-AI interactions~\cite{luo2025generative}. 

Our findings empirically extend prior discussions on the effectiveness of diachronic narrative on preservation awareness~\cite{Fu2024,lc2023speculative}, which could inform future interactive systems for educational and preservation purpose.

\subsection{Lowering Barriers to Interpretation and Fostering Cultural Belonging.}

LLMs can effectively embody expert roles over extended interactions~\cite{10.1145/3706598.3713109,10.1145/3746027.3754593,Trichopoulos2025chatbot}. This integration represents a novel approach to heritage interpretation, moving beyond static presentation models~\cite{Trichopoulos2025chatbot,xu2024shuangta}. Building on the design goals, we developed the LLM-based conversational agent. 

Through this agent, participants could access on-demand contextual explanations that expanded their understanding beyond surface-level facts, enabling richer and more multidimensional interpretations without requiring prior specialist knowledge. This design choice effectively lowered the barrier to entering the cultural-heritage discourse, allowing a broader range of visitors to engage with the Kaiping Diaolou and to articulate their own ideas and questions. 

At the same time, the authenticity guardrails effectively mitigated LLM hallucination issues and reduced the cognitive load by filtering out historically implausible or culturally inappropriate content, enabling participants to focus on sense-making and storytelling rather than on prompt debugging. In doing so, the system supported participants in narrating their own experiences, memories, and interpretations in ways that remained anchored to the local heritage context.

Our findings suggest that combining guided interpretation with creative expression can strengthen participants’ sense of identity and belonging. Several participants reported feeling more connected to the Kaiping Diaolou and “my cultural roots” after using \textit{Gen-Diaolou}, indicating that integrated AI-assisted interactive systems may support not only knowledge acquisition but also reflection related to cultural identity and place-based belonging.

\subsection{Design Implications}

Building on our empirical findings, we outline five design implications for future human–AI collaborative systems that support cultural-heritage learning, reflection, and safeguarding for future studies.

\subsubsection{Develop adaptable authenticity guardrails for diverse heritage contexts.}
While our authenticity guardrails were tailored to the architectural and task constraints of the Kaiping Diaolou, future systems should offer configurable, multi-layered frameworks that can be adapted to other heritage domains, including intangible practices, cultural relics, and art or archaeological sites.

\subsubsection{Support multiple temporalities to scaffold diachronic understanding.}
Given that \textit{Gen-Diaolou} facilitated effective past–present–future imagination and reflection, similar diachronic strategies can be generalized to other heritage contexts. Future designs should enable users to explore the evolution, vulnerability, and potential futures of heritage across different historical trajectories and socio-cultural processes.

\subsubsection{Enable community-centered co-creation across stakeholder groups.}
Beyond engaging residents and visitors, future systems should include interfaces and workflows that enable experts, practitioners, and diaspora communities to collaboratively contribute narratives, interpretations, and preservation ideas, thereby supporting more inclusive forms of heritage meaning-making.

\subsubsection{Extend risk visualization toward educational and policy contexts.}
The risk scenarios in our study proved effective in fostering an emotional connection to heritage threats. We propose leveraging similar visualizations to support school curricula, community workshops, and municipal planning discussions, thereby making abstract risks and preservation tradeoffs concrete  and actionable, ultimately cultivating preservation awareness.

\subsubsection{Establish ethical pipelines for provenance, moderation, and governance.}
As GenAI outputs enter public heritage discourse, systems should implement transparent provenance tracking, culturally appropriate moderation, and co-governed review mechanisms to prevent misrepresentation, over-simplification, or unintended cultural harm.

\subsection{Limitations and Future Work}\label{FUTURE WORK}

The museum-based field study broadened recruitment beyond the pilot, but our findings still primarily reflect the perspectives of general visitors. Future work should purposefully recruit participants with professional or long-term preservation backgrounds (e.g., conservators, museum practitioners, heritage planners) to examine how their experiences, expectations, and design needs may differ.

Second, the quiz in the pilot study mainly captured short-term factual recall and did not fully reflect narrative understanding or cross-period reasoning. Although we refined the knowledge questionnaire and added delayed post-tests in the field study, longitudinal deployments are still needed to strengthen these instruments and to track how awareness, identity, and interpretive skills evolve over time. 

Finally, future research should explore how different scaffolds and constraints shape users’ interpretive agency, and how communities and experts negotiate acceptable ranges of more inclusive participation in AI-assisted co-creation systems for CH.

\section{CONCLUSION}

\textit{Gen-Diaolou} integrates integrated AI-assisted learning and creativity to enrich cultural-heritage engagement. Across our studies, we found that combining guided historical exploration with generative co-creation not only improves visitors’ factual knowledge, but also strengthens their personal connection to heritage and their sense of responsibility for its preservation. Our work illustrates how human–AI collaboration in museum settings can shift cultural education from a largely passive, speculative-oriented experience to an active, imaginative dialogue with both past and future. We hope this approach will inform future human–AI systems for digital heritage, inspiring designs that treat visitors not only as learners, but also as potential stewards and co-creators of cultural memory.


\begin{acks}
{\color{black}This work was supported by the Computational Media and Arts Lab\footnote{Computational Media and Arts. See\url{https://cma.hkust-gz.edu.cn/about-cma/}} and the Jiangmen Wuyi Museum of Overseas Chinese. We thank Prof.~Selia Tan Jinhua, Professor at Wuyi University and a researcher in overseas Chinese history and heritage conservation, for her valuable advice and support. We also acknowledge support from The Hong Kong Wuyi Association through the \textit{GBA Future Internship Programme 2025}.}
\end{acks}




\bibliographystyle{ACM-Reference-Format}
\bibliography{sample-base}

\appendix

\section{Summary of Participant Demographics in the Formative, Pilot, and Field Studies}
\label{Qapp}
Table~\ref{PPFF} summarizes the demographics of participants across our formative study, pilot study, and museum field study.

\section{User Experience Questionnaire}
\label{sec:ux_questionnaire}

In the pilot study, we assessed participants’ subjective experience using the User Experience Questionnaire (UEQ) and workload using NASA-TLX; Table~\ref{tab:ueq_nasa_mapping} maps questionnaire items to their corresponding evaluation dimensions. In the museum field study, we measured preservation awareness, usability, and creativity support using CAI-CH, SUS, and CSI, respectively; Table~\ref{FSQ} maps these dimensions to the questionnaire items used in our study.

\begin{table*}[t]
  \centering
  \small
  \caption{Mapping of UEQ and NASA-TLX questionnaire items to evaluation dimensions.}
  \Description{Mapping of UEQ and NASA-TLX questionnaire items to evaluation dimensions.}
  \label{tab:ueq_nasa_mapping}
  \setlength{\tabcolsep}{6pt}
  \renewcommand{\arraystretch}{1.05}

  \begin{tabular}{@{}p{2.8cm} p{3.2cm} >{\raggedright\arraybackslash}p{8.8cm}@{}}
    \toprule
    \textbf{Instrument} & \textbf{Dimension} & \textbf{Questions} \\
    \midrule

    \multirow{5}{*}{\makecell[l]{\textbf{UEQ}}}
      & Perspicuity
      & The system was easy to learn and use. \newline
        The interface was clear and well-structured. \\
      & Efficiency
      & The system responded quickly to my actions. \newline
        The system supported me in achieving my goals. \\
      & Dependability
      & I felt confident while using the system. \newline
        The AI-generated results matched my expectations. \\
      & Stimulation
      & The system helped me express my creative ideas. \newline
        I enjoyed the overall experience of using the system. \\
      & Novelty
      & The system improved my understanding of CH. \newline
        I would like to use a similar system in the future. \\
    \midrule

    \multirow{6}{*}{\makecell[l]{\textbf{NASA-TLX}}}
      & Mental Demand
      & How much thinking, remembering, and attention was required for the task? \\
      & Physical Demand
      & How much physical activity or manual operation was required for the task? \\
      & Temporal Demand
      & How much time pressure did you feel during the task? \\
      & Performance
      & How satisfied are you with your performance in completing the task? \\
      & Effort
      & How hard did you have to work to achieve your level of performance? \\
      & Frustration Level
      & How impatient, stressed, or annoyed did you feel during the task? \\
    \bottomrule
  \end{tabular}
\end{table*}

\begin{table*}[t]
  \centering
  \small
  \caption{Mapping of CAI-CH, SUS, and CSI dimensions to their corresponding questionnaire items.}
  \Description{Mapping of CAI-CH, SUS, and CSI dimensions to their corresponding questionnaire items.}
  \label{FSQ}

  \setlength{\tabcolsep}{3pt}      
  \renewcommand{\arraystretch}{1.05}

  \begin{tabular}{@{}p{1.3cm} p{2.9cm} >{\raggedright\arraybackslash}p{11cm}@{}}
    \toprule
    \textbf{Instrument} & \textbf{Dimension} & \textbf{Questions} \\
    \midrule

    \multirow{5}{*}{\makecell[l]{\textbf{CAI-CH}}}
      & Value recognition
      & I can clearly recognize the historical, cultural, and social values embodied in Kaiping Diaolou CH.
        Kaiping Diaolou CH deserves protection because it represents irreplaceable values for society. \\
      & Heritage identity \& belonging
      & Kaiping Diaolou CH helps strengthen my sense of identity with the local community or culture.
        When I engage with Kaiping Diaolou, I feel a stronger sense of belonging to my cultural roots. \\
      & Intangible heritage awareness
      & I believe safeguarding intangible CH (e.g., oral traditions, rituals, and festive events) is as important as protecting physical sites.
        Participating in or observing traditional cultural practices related to Kaiping Diaolou increases my awareness of the need for heritage preservation. \\
      & Participation willingness
      & I am willing to engage in heritage-related activities (e.g., volunteer work, community events, educational programs).
        I believe that active public participation is essential for successful CH preservation. \\
      & Cultural sustainability
      & I feel a personal responsibility to ensure that CH is preserved for future generations.
        I am concerned that failure to protect CH today will cause irreversible losses for future society. \\
    \midrule

    \multirow{6}{*}{\makecell[l]{\textbf{SUS}}}
      & Easy to use
      & I thought the system was easy to use. \\
      & Functions
      & I found the various functions in this system were well integrated. \\
      & Quick to learn
      & I would imagine that most people would learn to use this system very quickly. \\
      & Learning curve
      & I needed to learn a lot of things before I could get going with this system. \\
      & Frequency
      & I think that I would like to use this system frequently. \\
      & Confidence
      & I felt very confident using the system. \\
    \midrule

    \multirow{6}{*}{\makecell[l]{\textbf{CSI}}}
      & Enjoyment
      & I enjoyed using this system. The experience was fun and engaging. I felt positive while using the system. \\
      & Exploration
      & The system helped me to track different ideas, outcomes, or possibilities.
        The system encouraged me to try new things.
        It was easy to experiment with alternatives.
        I discovered new possibilities. \\
      & Collaboration
      & I was able to share or discuss my ideas with others.
        It was easy to build on others’ ideas.
        The system supported collaboration. \\
      & Results worth effort
      & I could communicate with others about my work.
        The results were worth the effort required.
        The quality of outcomes justified the time spent.
        The system helped me produce valuable results. \\
      & Expressiveness
      & The system allowed me to be very expressive.
        The system supported my personal style.
        I could create exactly what I intended.
        I could convey my ideas effectively.
        I felt in control of the final result. \\
      & Immersion
      & I lost track of time while using it.
        I was deeply engaged in the activity.
        I felt absorbed in what I was doing.
        The interface did not distract me.
        My attention was fully captured. \\
    \bottomrule
  \end{tabular}
\end{table*}

\section{Task Performance and Iteration Patterns}
\label{TaskPerformance}

In the pilot study, we logged participants’ task performance in the \textit{GenAI Module}, including the number of generated images and prompt iterations for Task~1--3. Table~\ref{tab:aigc_task} reports the total counts per participant.

\section{Pilot Study Feedback and Iteration}
\label{FI}

Table~\ref{tab:iteration} summarizes key design implications and corresponding iteration suggestions derived from the pilot study. We used these findings to implement actionable improvements prior to the subsequent field study.


\begin{table*}[t]
\centering
\caption{Number of generated images and prompt iterations across \textit{GenAI Module} tasks (P1--P18).}
\Description{Number of generated images and prompt iterations across \textit{GenAI Module} tasks (P1--P18).}
\label{tab:aigc_task}
\small
\setlength{\tabcolsep}{3pt}
\renewcommand{\arraystretch}{1.05}
\begin{tabular}{|c|c|*{18}{c|}}
\hline
\textbf{Stage} &  & P1 & P2 & P3 & P4 & P5 & P6 & P7 & P8 & P9 & P10 & P11 & P12 & P13 & P14 & P15 & P16 & P17 & P18 \\
\hline
\multirow{2}{*}{Task 1} 
 & Generated Images  & 4 & 12 & 8 & 8 & 4 & 4 & 4 & 4 & 4 & 4 & 4 & 8 & 8 & 4 & 8 & 12 & 16 & 8 \\
\cline{2-20}
 & Prompt Iterations & 1 & 3 & 2 & 2 & 1 & 1 & 1 & 1 & 1 & 1 & 1 & 2 & 2 & 1 & 2 & 3 & 4 & 2 \\
\hline
\multirow{2}{*}{Task 2} 
 & Generated Images  & 8 & 4 & 12 & 4 & 4 & 4 & 4 & 8 & 4 & 4 & 4 & 4 & 8 & 4 & 4 & 8 & 4 & 8 \\
\cline{2-20}
 & Prompt Iterations & 2 & 1 & 3 & 1 & 1 & 1 & 1 & 2 & 1 & 1 & 1 & 1 & 2 & 1 & 1 & 2 & 1 & 2 \\
\hline
\multirow{2}{*}{Task 3} 
 & Generated Images  & 8 & 4 & 4 & 12 & 4 & 4 & 8 & 4 & 8 & 4 & 4 & 4 & 4 & 4 & 4 & 4 & 4 & 4 \\
\cline{2-20}
 & Prompt Iterations & 2 & 1 & 1 & 3 & 1 & 1 & 2 & 1 & 2 & 1 & 1 & 1 & 1 & 1 & 1 & 1 & 1 & 1 \\
\hline
\multicolumn{20}{|c|}{\textbf{Summary (M/SD)}} \\
\hline
 & Generated Images  & \multicolumn{18}{c|}{Task 1: 7.3 / 3.4 \quad Task 2: 6.0 / 2.5 \quad Task 3: 5.6 / 2.4} \\
\hline
 & Prompt Iterations & \multicolumn{18}{c|}{Task 1: 1.8 / 0.9 \quad Task 2: 1.6 / 0.7 \quad Task 3: 1.3 / 0.7} \\
\hline
\end{tabular}
\end{table*}

\begin{table*}[t]
\centering
\caption{System iteration suggestions based on findings from the post-pilot study.}
\Description{System iteration suggestions based on findings from the post-pilot study.}
\label{tab:iteration}

\small
\setlength{\tabcolsep}{4pt}
\renewcommand{\arraystretch}{1.08}

\begin{tabular}{@{}p{2.8cm} >{\raggedright\arraybackslash}p{6.8cm} >{\raggedright\arraybackslash}p{6.8cm}@{}}
\toprule
\textbf{Area} & \textbf{Iteration suggestion} & \textbf{Motivation from findings} \\
\midrule

\textbf{Interaction Flow} &
Clarify task goals with short task tips; optionally provide site-specific background soundscapes to enhance immersion during exploration and creation. &
Some participants felt unsure about what each task expected and how to move from knowledge exploration to creative generation; some explicitly mentioned that background sound could improve the overall atmosphere ($P7$, $P12$, $P13$). \\
\addlinespace

\textbf{Knowledge Module} &
Add richer narrative content (e.g., family histories, key events) with curated narrative pathways, and incorporate cultural-heritage preservation case studies into the \textit{Speculative Futures} theme to ground users’ future-oriented ideas. &
Some participants appreciated the structured taxonomy but asked for richer, story-based content and tighter linkage between heritage knowledge and GenAI prompts ($P5$, $P9$). \\
\addlinespace

\textbf{GenAI Module} &
Provide task-specific prompt templates and modular slots (e.g., component, pattern); show both a structured and a natural-language view of the scaffolded prompt. &
Some users struggled to translate ideas into effective prompts and wanted clearer guidance about how their inputs were being rephrased and expanded ($P12$, $P16$). \\
\addlinespace

\textbf{Authenticity Guardrails} &
When an input triggers a guardrail, show a pop-up explaining the block and offering historically plausible alternatives. &
Users occasionally interpreted guardrail interventions as model errors and expressed a desire to balance historical fidelity with imaginative exploration ($P1$, $P2$, $P4$, $P16$, $P18$). \\
\addlinespace

\textbf{Output \& Presentation} &
Add comparison views and exportable ``exhibit cards,'' and store user-facing views with a system-assigned creation ID so users can sequentially review their past creations. &
Participants wanted better support for reflecting on design changes over time and for sharing or exhibiting their favorite outcomes ($P3$, $P8$, $P13$). \\
\bottomrule
\end{tabular}
\end{table*}

\section{Authenticity Guardrails System Prompts}\label{appendix:guardrails}

We configure the authenticity guardrails with fidelity-first defaults for Historical Reconstruction and more exploration-friendly settings for themes of risk, challenges, and protection (see Table~\ref{ttier2} and Table~\ref{tab:guardrails-comparison}).

Table~\ref{ttier2} specifies the Tier~2 tag vocabulary across eight categories: \textit{Viewpoint}, \textit{Time of Day}, \textit{People}, \textit{Building Function}, \textit{Architectural Style}, \textit{Window Features}, \textit{Decorative Patterns}, and \textit{Rendering Style}. These user-selectable tags are treated as hard requirements and directly guide prompt assembly, ensuring that generated images align with the selected attributes.

Table~\ref{tab:guardrails-comparison} details how guardrail constraints adapt across the three GenAI module tasks. \textit{Task~1 (Historical Reconstruction)} enforces strict 1930s period accuracy with minimal tolerance for anachronisms; \textit{Task~2 (Risk Estimation)} relaxes temporal constraints to allow present or near-future scenarios while maintaining architectural integrity; and \textit{Task~3 (Future Preservation)} encourages creative speculation for preservation planning while preserving the recognizable Diaolou form. This tiered approach operationalizes domain knowledge as context-sensitive constraints that balance historical fidelity with task-appropriate creative freedom.

In the \textit{Historical Reconstruction} theme, historically inaccurate content (e.g., \textit{``glass curtain wall,'' ``futuristic skyscraper''}) is normalized under Tier~1 and rewritten as period-appropriate alternatives (e.g., \textit{masonry façades, iron-grille windows}); culturally inappropriate motifs are replaced with authentic Chinese references.

After validation, the system enriches the prompt with Diaolou-specific vocabulary from the knowledge base (e.g., brick-and-stone façades, roof forms, parapets, defensive ``swallow's-nest'' loopholes) to increase architectural fidelity.

\begin{table*}[t]
  \centering
  \caption{Tier~2---Tag specifications for prompt assembly across GenAI module tasks.}
  \Description{Tier~2---Tag specifications for prompt assembly across GenAI module tasks.}
  \label{ttier2}

  {\ttfamily\small
  \setlength{\tabcolsep}{5pt}
  \renewcommand{\arraystretch}{1.05}

  \begin{tabular}{@{}p{3.0cm} >{\raggedright\arraybackslash}p{14cm}@{}}
    \toprule
    \textbf{Tag Category} & \textbf{Specifications} \\
    \midrule

    \textbf{Viewpoint} &
    Distant view (100--200\,m, 20--30\% frame coverage); Medium view (25--45\,m, 65--75\% frame, eye-level); Close-up (10--20\,m, 80--90\% frame). \\
    \addlinespace

    \textbf{Time of Day} &
    Morning (6--9\,AM, low-angle golden light); Afternoon (12--4\,PM, overhead bright light); Evening (5--7\,PM, warm sunset light). \\
    \addlinespace

    \textbf{People} &
    None (uninhabited scene); Single (one period-appropriate figure); Multiple (3--8 individuals in traditional attire). \\
    \addlinespace

    \textbf{Building Function} &
    Defense-focused (watchtower, gun ports); Flood protection (elevated foundation); Residential (domestic life scenes). \\
    \addlinespace

    \textbf{Architectural Style} &
    Romanesque; Baroque; Byzantine; Indo-British; Neoclassical. \\
    \addlinespace

    \textbf{Window Features} &
    Yanhu (Baroque-style); Changhu (Neoclassical); Liuhu (Romanesque); Dense grid pattern; Linhu (Byzantine). \\
    \addlinespace

    \textbf{Decorative Patterns} &
    Plant motifs; Animal patterns; Geometric designs (interior views only). \\
    \addlinespace

    \textbf{Rendering Style} &
    Photorealistic; Oil painting (classical European); Ink wash painting; Gongbi (traditional Chinese meticulous painting);
    Impressionist; Pointillist. \\
    \bottomrule
  \end{tabular}
  }
\end{table*}

\begin{table*}[t]
  \centering
  \caption{Task-specific authenticity guardrail constraints across GenAI module tasks.}
  \Description{Task-specific authenticity guardrail constraints across GenAI module tasks.}
  \label{tab:guardrails-comparison}

  {\ttfamily\small
  \setlength{\tabcolsep}{4pt} 
  \renewcommand{\arraystretch}{1.05}

  \begin{tabular}{@{}%
    >{\raggedright\arraybackslash}p{3.0cm}%
    >{\raggedright\arraybackslash}p{4.5cm}%
    >{\raggedright\arraybackslash}p{4.5cm}%
    >{\raggedright\arraybackslash}p{4.5cm}@{}}
    \toprule
    \textbf{Constraint Aspect} & \textbf{Task 1} & \textbf{Task 2} & \textbf{Task 3} \\
    \midrule

    \makecell[l]{\textbf{Tier 1.}\\\textbf{Temporal Constraint}} &
    Strict 1930s only; all elements must conform to the historical period. &
    Present or near-future allowed to depict realistic threats. &
    Speculative future encouraged for preservation scenarios. \\
    \addlinespace

    \makecell[l]{\textbf{Tier 1.}\\\textbf{Architectural Integrity}} &
    Fundamental structure (form, proportions, façade, roofline, window positions) MUST remain unchanged. &
    Fundamental structure MUST remain unchanged; can show deterioration. &
    Recognizable Diaolou form MUST remain intact; allows future adaptive reuse. \\
    \addlinespace

    \makecell[l]{\textbf{Tier 1.}\\\textbf{Cultural Context}} &
    Kaiping, Guangdong, China setting mandatory; all figures must be Chinese with era-appropriate culture; strictly avoid modern anachronisms. &
    Kaiping, Guangdong setting with authentic cultural elements; anachronisms allowed for risk visualization. &
    Chinese cultural context maintained; future community engagement scenarios permitted. \\
    \addlinespace

    \makecell[l]{\textbf{Tier 3.}\\\textbf{Validation Hierarchy}} &
    If user idea conflicts with tags, tags take precedence; if input violates 1930s rules, automatically normalized; non-conforming descriptions rewritten or removed. &
    If user idea conflicts with tags, tags take precedence; temporal constraints relaxed; architectural integrity strictly enforced. &
    If user idea conflicts with tags, tags take precedence; temporal constraints fully relaxed; architectural form still enforced; creative speculation permitted. \\
    \addlinespace

    \textbf{Allowed Content} &
    1930s Kaiping context only. For interior views: preserve architectural structure (walls, layout, windows, doors, ceiling, floor), modify ONLY decorative elements, exclude furniture and people, use period Chinese decorative arts. &
    Risk-related content permitted: water damage, weathering, structural deterioration, foundation issues, visitor impact, environmental threats. &
    Preservation content encouraged: future scenarios, community participation, sustainable development, heritage protection strategies. \\
    \addlinespace

    \textbf{Example Normalization} &
    \textit{Input:} ``futuristic glass curtain wall'' $\rightarrow$ \textit{Output:} ``masonry façades with iron-grille windows''; \textit{Input:} ``tanks and armored vehicles'' $\rightarrow$ \textit{Output:} removed or replaced with period-appropriate elements. &
    \textit{Input:} ``Diaolou completely demolished'' $\rightarrow$ \textit{Output:} ``Diaolou with visible structural cracks and weathering''. &
    \textit{Input:} ``transform into sci-fi spaceship'' $\rightarrow$ \textit{Output:} ``heritage-themed cultural center with interactive projection mapping, preserving original tower form''. \\
    \bottomrule
  \end{tabular}
  }
\end{table*}

\begin{table*}[t]
  \centering
  \scriptsize
  \setlength{\tabcolsep}{3pt}
  \renewcommand{\arraystretch}{1.00}

  \caption{Participant demographics across the formative, pilot, and field studies. Columns include session, ID, gender, background, education, age, self-reported GenAI proficiency for idea generation, and Diaolou knowledge pre-score.}
  \Description{Participant demographics across the formative, pilot, and field studies. Columns include session, ID, gender, background, education, age, self-reported GenAI proficiency for idea generation, and Diaolou knowledge pre-score.}
  \label{PPFF}

  \begin{tabular}{@{}
    >{\centering\arraybackslash}m{2cm}
    >{\centering\arraybackslash}p{1.55cm}
    >{\centering\arraybackslash}p{1.00cm}
    >{\raggedright\arraybackslash}p{3cm}
    >{\centering\arraybackslash}p{1.55cm}
    >{\centering\arraybackslash}p{1.10cm}
    >{\centering\arraybackslash}p{1.30cm}
    >{\centering\arraybackslash}p{1.10cm}@{}}
    \toprule
    \textbf{Session} & \textbf{ID} & \textbf{Gender} & \textbf{Background} &
    \textbf{Education} & \textbf{Age} & \textbf{GenAI Exp.\textsuperscript{1}} & \textbf{Pre Score\textsuperscript{2}} \\
    \midrule

    \multirow{14}{*}{\makecell[c]{\textbf{Formative}\\\textbf{Study}}}
      & E1  & Female & Heritage scholar                & PhD         & 56 & 3 & Expert \\
      & E2  & Female & Local museum head               & Master’s    & 46 & 3 & Expert \\
      \cmidrule(lr){2-8}
      & C1  & Female & Information design              & Bachelor’s  & 19 & 3 & / \\
      & C2  & Male   & Computer science                & Bachelor’s  & 21 & 4 & / \\
      & C3  & Female & Computer science                & PhD         & 27 & 3 & / \\
      & C4  & Male   & History                         & Bachelor’s  & 19 & 2 & / \\
      & C5  & Female & Information design              & Master’s    & 22 & 4 & / \\
      & C6  & Female & Electronic engineering          & Bachelor’s  & 23 & 3 & / \\
      \cmidrule(lr){2-8}
      & C7  & Female & Data science                    & Master’s    & 24 & 4 & / \\
      & C8  & Male   & Media art                       & PhD         & 25 & 5 & / \\
      & C9  & Female & Design                          & Master’s    & 27 & 4 & / \\
      & C10 & Male   & Public policy                   & PhD         & 25 & 3 & / \\
      & C11 & Male   & Artificial intelligence         & Bachelor’s  & 20 & 1 & / \\
      & C12 & Male   & Bioscience                      & PhD         & 28 & 1 & / \\
    \midrule

    \multirow{18}{*}{\makecell[c]{\textbf{Pilot}\\\textbf{Study}}}
      & P1  & Male   & Design                          & PhD         & 29 & 3 & 2 \\
      & P2  & Female & Public policy                   & Master’s    & 21 & 2 & 12 \\
      & P3  & Male   & Bioengineering                  & Bachelor’s  & 22 & 1 & 4 \\
      & P4  & Male   & Film-making                     & Master’s    & 23 & 3 & 12 \\
      & P5  & Male   & Media art                       & PhD         & 30 & 2 & 3 \\
      & P6  & Female & Computer science                & PhD         & 26 & 2 & 10 \\
      & P7  & Female & Engineering                     & Master’s    & 24 & 2 & 5 \\
      & P8  & Male   & Bioengineering                  & PhD         & 28 & 5 & 13 \\
      & P9  & Female & Computer science                & PhD         & 24 & 3 & 8 \\
      & P10 & Male   & Intelligent manufacturing       & Bachelor’s  & 20 & 1 & 11 \\
      & P11 & Male   & Artificial intelligence         & Bachelor’s  & 19 & 3 & 12 \\
      & P12 & Male   & Public policy                   & PhD         & 23 & 2 & 10 \\
      & P13 & Female & Computer science                & PhD         & 29 & 3 & 1 \\
      & P14 & Male   & Artificial intelligence         & Bachelor’s  & 19 & 3 & 7 \\
      & P15 & Male   & Public policy                   & Bachelor’s  & 23 & 1 & 9 \\
      & P16 & Female & Design                          & PhD         & 31 & 2 & 7 \\
      & P17 & Male   & Computer science                & Bachelor’s  & 21 & 3 & 4 \\
      & P18 & Female & Industrial design               & Bachelor’s  & 23 & 1 & 10 \\
    \midrule

    \multirow{13}{*}{\makecell[c]{\textbf{Field Study}\\\textbf{(Base)}}}
      & F1  & Female & Local resident                  & Bachelor’s        & 20 & 3 & 10 \\
      & F2  & Female & Local resident                  & Vocational school & 21 & 3 & 9 \\
      & F3  & Female & In-province visitor             & Bachelor’s        & 35 & 3 & 15 \\
      & F4  & Female & Out-of-province visitor         & Bachelor’s        & 35 & 4 & 9 \\
      & F5  & Female & In-province visitor             & Vocational school & 20 & 4 & 15 \\
      & F6  & Female & Local resident                  & Bachelor’s        & 33 & 3 & 12 \\
      & F7  & Female & Local resident                  & Bachelor’s        & 28 & 4 & 11 \\
      & F8  & Female & Local resident                  & Bachelor’s        & 22 & 3 & 16 \\
      & F9  & Female & Out-of-province tourist         & Bachelor’s        & 25 & 5 & 13 \\
      & F10 & Female & Out-of-province tourist         & Bachelor’s        & 37 & 4 & 12 \\
      & F11 & Female & Out-of-province tourist         & Bachelor’s        & 28 & 1 & 11 \\
      & F12 & Female & Local resident                  & Master’s          & 23 & 4 & 14 \\
      & F13 & Male   & Out-of-province tourist         & Bachelor’s        & 39 & 4 & 14 \\
    \midrule

    \multirow{13}{*}{\makecell[c]{\textbf{Field Study}\\\textbf{(Learn+GenAI)}}}
      & F14 & Female & Out-of-province visitor         & Bachelor’s        & 43 & 4 & 12 \\
      & F15 & Female & Local resident                  & Master’s          & 25 & 4 & 11 \\
      & F16 & Male   & Local resident                  & Bachelor’s        & 42 & 3 & 15 \\
      & F17 & Female & Out-of-province tourist         & Bachelor’s        & 27 & 3 & 8 \\
      & F18 & Female & Local resident                  & Bachelor’s        & 41 & 3 & 15 \\
      & F19 & Female & Local resident                  & Bachelor’s        & 30 & 4 & 13 \\
      & F20 & Female & Local resident                  & Bachelor’s        & 43 & 2 & 17 \\
      & F21 & Female & In-province visitor             & Master’s          & 35 & 1 & 11 \\
      & F22 & Male   & Out-of-province tourist         & Vocational school & 24 & 2 & 12 \\
      & F23 & Female & In-province visitor             & Bachelor’s        & 23 & 5 & 16 \\
      & F24 & Female & Out-of-province tourist         & Bachelor’s        & 23 & 4 & 9 \\
      & F25 & Female & Local resident visitor          & Master’s          & 40 & 3 & 14 \\
      & F26 & Female & Out-of-province tourist         & Master’s          & 20 & 3 & 16 \\
    \bottomrule
  \end{tabular}

  \vspace{0.3em}
  \begin{minipage}{\textwidth}
    \footnotesize
    \textsuperscript{1} Self-reported proficiency with GenAI for idea generation: 1 (Novice) to 5 (Expert).\\
    \textsuperscript{2} Pre score denotes the baseline Diaolou knowledge score tested before using the system/visit: pilot study (max = 15); field study (max = 20). Formative-study experts were not administered the pre-test.
  \end{minipage}

\end{table*}

\end{document}